\documentclass[aps,prl,reprint,superscriptaddress,longbibliography]{revtex4-1}

\usepackage{amsmath,amssymb,graphicx,bm}
\usepackage{xcolor}
\usepackage[colorlinks=true,urlcolor=blue,linkcolor=blue,citecolor=blue,bookmarks=false]{hyperref}
\usepackage{changes}

\begin{document}

\title{Resonant elastic X-ray scattering of antiferromagnetic superstructures in EuPtSi$_{\text{3}}$}

\author{Wolfgang Simeth}
\address{Physik-Department, Technische Universit\"at M\"unchen, 85748 Garching, Germany}
\address{Laboratory for Neutron and Muon Instrumentation, Paul Scherrer Institute, Villigen PSI, Switzerland}

\author{Andreas Bauer}
\address{Physik-Department, Technische Universit\"at M\"unchen, 85748 Garching, Germany}
\address{Zentrum f\"ur QuantumEngineering (ZQE), Technische Universit\"at M\"unchen, D-85748 Garching, Germany}

\author{Christian Franz}
\address{Physik-Department, Technische Universit\"at M\"unchen, 85748 Garching, Germany}
\address{J\"ulich Centre for Neutron Science (JCNS) at Heinz Maier-Leibnitz Zentrum (MLZ), D-85748 Garching, Germany}

\author{Aisha Aqeel} 
\address{Physik-Department, Technische Universit\"at M\"unchen, 85748 Garching, Germany}
\address{Munich Center for Quantum Science and Technology (MCQST), Technische Universit\"at M\"unchen, D-85748 Garching, Germany}

\author{Pablo J. Bereciartua Perez}
\address{Deutsches Elektronen-Synchrotron (DESY), D-22607 Hamburg, Germany}

\author{Jennifer A. Sears} 
\address{Deutsches Elektronen-Synchrotron (DESY), D-22607 Hamburg, Germany}

\author{Sonia Francoual} 
\address{Deutsches Elektronen-Synchrotron (DESY), D-22607 Hamburg, Germany}

\author{Christian H. Back} 
\address{Physik-Department, Technische Universit\"at M\"unchen, 85748 Garching, Germany}
\address{Zentrum f\"ur QuantumEngineering (ZQE), Technische Universit\"at M\"unchen, D-85748 Garching, Germany}
\address{Munich Center for Quantum Science and Technology (MCQST), Technische Universit\"at M\"unchen, D-85748 Garching, Germany}

\author{Christian Pfleiderer}
\address{Physik-Department, Technische Universit\"at M\"unchen, 85748 Garching, Germany}
\address{Zentrum f\"ur QuantumEngineering (ZQE), Technische Universit\"at M\"unchen, D-85748 Garching, Germany}
\address{Munich Center for Quantum Science and Technology (MCQST), Technische Universit\"at M\"unchen, D-85748 Garching, Germany}

\date{\today}

\begin{abstract}
We report resonant elastic X-ray scattering (REXS) of long-range magnetic order in EuPtSi$_{\text{3}}$, combining different scattering geometries with full linear polarization analysis to unambiguously identify magnetic scattering contributions. At low temperatures, EuPtSi$_{\text{3}}$ stabilizes type A antiferromagnetism featuring various long-wavelength modulations. For magnetic fields applied in the hard magnetic basal plane, well-defined regimes of cycloidal, conical, and fan-like superstructures may be distinguished that encompass a pocket of commensurate type A order without superstructure. For magnetic field applied along the easy axis, the phase diagram comprises the cycloidal and conical superstructures only. Highlighting the power of polarized REXS, our results reveal a combination of magnetic phases that suggest a highly unusual competition between antiferromagnetic exchange interactions with Dzyaloshinsky--Moriya spin--orbit coupling of similar strength.
\end{abstract}

\maketitle

In recent years great efforts have been made to identify magnetic superstructures in bulk materials, thin films, and nano-scaled systems~\cite{1982_Bak_RepProgPhys, 1984_Izyumov_SovPhysUsp, 1990_Cummins_PhysRep, 2013_Nagaosa_NatNanotechnol, 2020_Back_JPhysD}. In systems comprising ferromagnetic exchange with Dzyaloshinsky--Moriya (DM) spin--orbit coupling~\cite{1958_Dzyaloshinsky_JPhysChemSolids, 1960_Moriya_PhysRev} major discoveries include long-wavelength incommensurate modulations~\cite{1976_Ishikawa_SolidStateCommun, 1976_Motoya_SolidStateCommun, 1981_Beille_JPhysF, 1993_Lebech_Book}, solitonic structures~\cite{2012_Togawa_PhysRevLett}, and topologically nontrivial order such as skyrmion lattices~\cite{2009_Muhlbauer_Science, 2010_Yu_Nature, 2012_Kanazawa_PhysRevB, 2016_Kanazawa_NewJPhys, 2016_Bauer_Book, 2017_Bauer_PhysRevB, 2018_Chacon_NatPhys}. While these modulated states under applied magnetic field may feature transitions between different superstructures, they collapse at a well-defined transition into a field-polarized state~\cite{1970_Lundgren_PhysScr, 1975_Bloch_PhysLettA, 2010_Bauer_PhysRevB}. In comparison, less is known about materials comprising antiferromagnetic exchange with DM interactions, as the mere number of possible modulated structures is much larger. Representing the perhaps most general condition, an unresolved question concerns possible magnetic order in the presence of antiferromagnetic exchange and DM interactions of similar strength. 

Focusing on magnetic ions such as Eu$^{2+}$ or Gd$^{3+}$, in which quenched orbital momentum gives way to almost unconstrained spin degrees of freedom, a rich variety of antiferromagnetic states has attracted great interest. Topical examples include incommensurate antiferromagnetism and a large topological Hall effect in EuGa$_2$Al$_2$ and EuAl$_4$~\cite{2018_Stavinoha_PhysRevB, 2021_Shang_PhysRevB, 2022_Meier_PhysRevB, 2022_Zhu_PhysRevB, 2022_Moya_PhysRevMater, 2022_Takagi_NatCommun}, complex antiferromagnetism in GdRh$_2$Si$_2$, skyrmion lattice order in GdRu$_2$Si$_2$~\cite{2020_Khanh_NatNanotechnol, 2020_Yasui_NatCommun} and Gd$_2$PdSi$_3$~\cite{2019_Kurumaji_Science, 2019_Hirschberger_NatCommun}, colossal magnetoresistance in antiferromagnetic Zintl compounds such as Eu$X_{2}Y_{2}$ ($X=\text{Cd, In}$, and $Y=\text{As, Sb, P}$)~\cite{2006_Jiang_ChemMater, 2008_Goforth_InorgChem, 2010_Cao_PhysRevB, 2014_Rosa_JMagnMagnMater, 2018_Ghimire_NatCommun, 2018_Rahn_PhysRevB, 2018_Soh_PhysRevB, 2018_Takahashi_SciAdv, 2020_Asaba_PhysRevB, 2021_Gaudet_NatMater, 2021_Wang_AdvMater}, as well as magnetic order and superconductivity in Eu$X_2$As$_2$ ($X=\text{Fe, Ni, Cr, Co}$) and related compounds~\cite{2013_Seiro_JPhysCondensMatter, 2019_Jin_PhysRevB, 2021_Xie_PhysRevB}. These systems, however, lack global DM interactions in their centrosymmetric crystal structures. This is contrasted by the observation of magnetic superstructures, superconductivity, and quantum criticality in Eu$TX_3$, where $T=\text{Pt, Pd, Ni, Rh, Co, Ir}$ and $X=\text{Si, Ge, Sn, Ga}$, most of which lack inversion symmetry~\cite{2005_Kimura_PhysRevLett, 2006_Sugitani_JPhysSocJpn, 2012_Kumar_JPhysCondensMatter, 2012_Kaczorowski_SolidStateCommun, 2014_Albedah_JAlloyCompd, 2014_Maurya_JPhysCondensMatter, 2015_Bednarchuk_ActaPhysPolA, 2015_Bednarchuk_JAlloyCompd, 2015_Bednarchuk_JAlloyCompda, 2016_Maurya_JMagnMagnMater, 2016_Fabreges_PhysRevB, 2022_Matsumura_JPhysSocJpn, 2022_Rai_JPhysCondensMatter}. 

\begin{figure}
\centering
\includegraphics{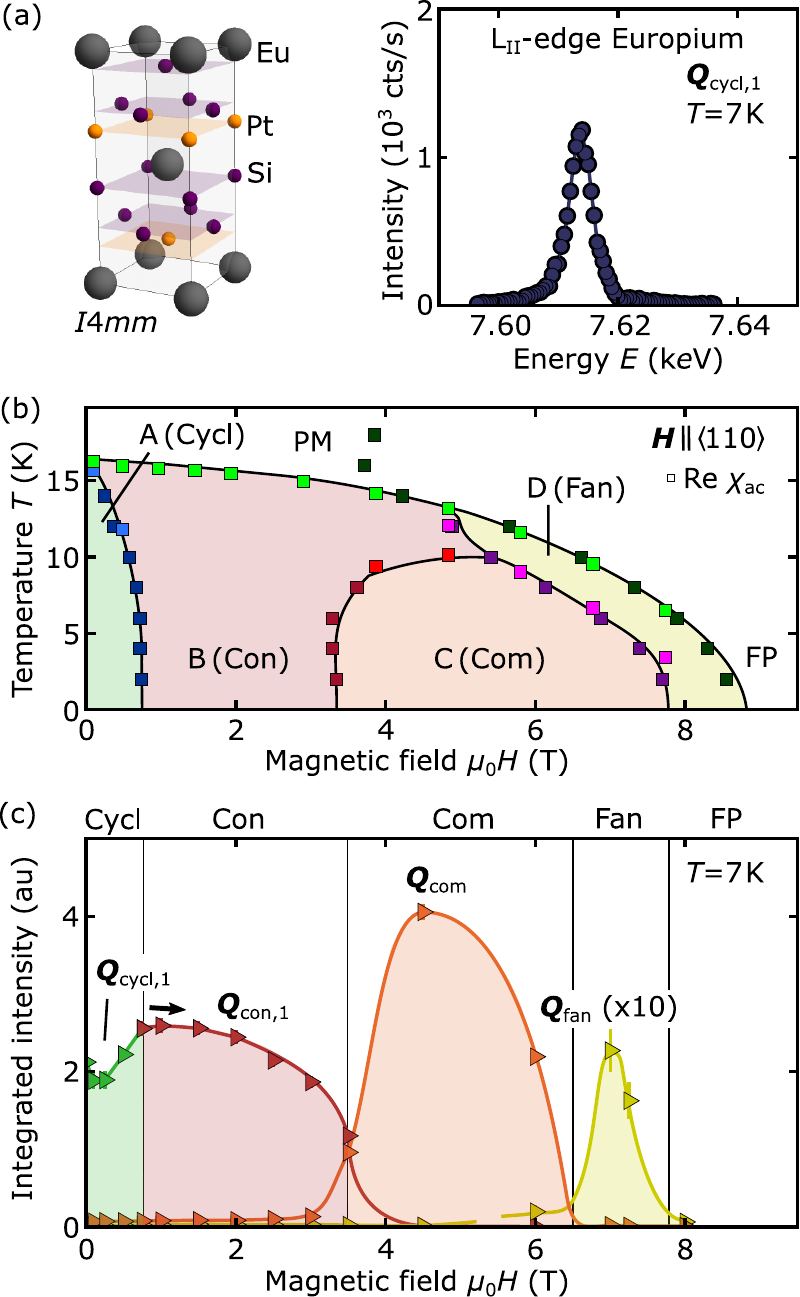}
\caption{Crystallographic and magnetic properties of EuPtSi$_3$. (a)~Tetragonal unit cell of EuPtSi$_{3}$, space group $I4mm$, and resonant enhancement of the magnetic Bragg intensity when tuning the incident photon energy across the L$_{\mathrm{II}}$ edge of europium, characteristic of magnetism predominantly carried by Eu$^{2+}$ moments. (b)~Magnetic phase diagram for field parallel to $[\bar{1}10]$, as inferred from susceptibility, $\mathrm{Re}\,\chi_{\mathrm{ac}}$~\cite{2022_Bauer_PhysRevMater}. Antiferromagnetic phases with cycloidal (Cycl, green), conical (Con, red), commensurate (Com, orange), and fan-like (Fan, yellow) order as well as paramagnetic~(PM) and field-polarized~(FP) regimes may be distinguished. (c)~Field dependence of the REXS intensities at Bragg positions characteristic of the different magnetic phases.}
\label{figure1}
\end{figure}

For our study, we selected EuPtSi$_{3}$, which crystallizes in the noncentrosymmetric tetragonal BaNiSn$_{3}$ structure (space group $I4mm$), shown in Fig.~\ref{figure1}(a)~\cite{2010_Kumar_PhysRevB}. Measurements of the bulk properties established the characteristics of antiferromagnetic order of localized Eu$^{2+}$ moments below a transition temperature $T_{\mathrm{N}} = 17\,\mathrm{K}$~\cite{2022_Bauer_PhysRevMater}. Depending on field direction, up to four different phase pockets, denoted A through D, were identified, as shown in Fig.~\ref{figure1}(b) for field parallel $\langle110\rangle$. For the point group symmetry of EuPtSi$_{3}$, DM vectors $\bm{D}_{ij}$ are permitted that support the formation of superlattice structures with N{\'e}el-type twisting including antiferromagnetic N{\'e}el skyrmions~\cite{1989_Bogdanov_SovPhysJETP, 2020_Back_JPhysD}. Preliminary neutron scattering suggested some form of superlattice modulation with a wavelength of about $100~\text{\AA}$ at zero field, however, without information on the nature of the underlying antiferromagnetism~\cite{2022_Bauer_PhysRevMater}.

Experimentally, the unambiguous determination of complex antiferromagnetic spin structures is especially demanding, requiring scattering techniques with high momentum resolution at large momentum transfers, the possibility to obtain element-specific information, and to separate spin and orbital degrees of freedom. Moreover, in many cases tiny sample volumes are available only, e.g., in the form of thin films, nano-scale systems, or bulk samples of highest purity. While neutron scattering has become indispensable in studies of magnetic structures, it cannot meet these general requirements, not to mention prohibitively strong neutron absorption in elements such as Eu, Gd, or Cd. In contrast, seminal studies of comparatively simple magnetic structures in selected rare-earth compounds have demonstrated the capability of resonant elastic X-ray scattering (REXS) with full linear polarization analysis (FLPA) to overcome these challenges~\cite{2007_Mazzoli_PhysRevB, 2009_Hatton_JMagnMagnMater, 2012_Shukla_PhysRevB, 2016_Zhang_PhysRevB, 2018_Rahn_PhysRevB}.

Using REXS, we determined the four antiferromagnetically ordered phases of EuPtSi$_{3}$. As a main result, we find that all antiferromagnetic phases represent variations of the same type A antiferromagnetism, where the tetragonal $[001]$ axis is the easy magnetic direction. Combining different scattering geometries and FLPA, we identify long-wavelength cycloidal, conical, and fan-like superstructures, consistent with the DM vectors expected in space group $I4mm$~\cite{2022_Bauer_PhysRevMater}. At intermediate fields, the conical and fan-like superstructure encompass a phase pocket of pure type A antiferromagnetism without superstructure, reflecting antiferromagnetic exchange and DM interactions of similar strength. For field along $[001]$, only the phases with cycloidal and conical superstructures are stabilized.

For the REXS experiments, a polished single-crystal cube with an edge length of 2~mm was used as prepared from an ingot grown by means of the optical floating-zone technique~\cite{SOI, 2016_Bauer_RevSciInstrum}. The same sample was also used for the study of the magnetization, ac susceptibility, and specific heat reported in Ref.~\cite{2022_Bauer_PhysRevMater}. REXS was carried out in the second experimental hutch EH2 of beamline P09 at the synchrotron source PETRA~III~\cite{2013_Strempfer_JSynchrotronRad}. Hard X-rays at an incident photon energy of 7.61~keV were used close to the L$_{\mathrm{II}}$ edge of europium, cf.\ Fig.~\ref{figure1}(a), where the magnetic cross-section is dominated by electric dipole transitions~\cite{1996_Hill_ActaCrystallogrA}. The X-ray diffraction was carried out in a horizontal scattering geometry and the polarization of the scattered beam was analyzed using PG006 as the analyzer crystal~\cite{2013_Francoual_JPhysConfSer, SOI}, permitting a clean polarization analysis. Magnetic structure refinement by means of a FLPA was carried out using a double phase-retarder~\cite{2013_Francoual_JPhysConfSer} in combination with the analyzer. In this setup, the polarization plane of the incident beam was rotated rather than the sample, which permits to avoid parasitic mixing due to slightly different beam spot positions on the sample as well as differences of angular positions of the diffractometer. Further information on the sample alignment, the mathematical description of polarized REXS, and the magnetic structure determination may be found in the supplementary information~\cite{SOI}.

\begin{figure}
\centering
\includegraphics{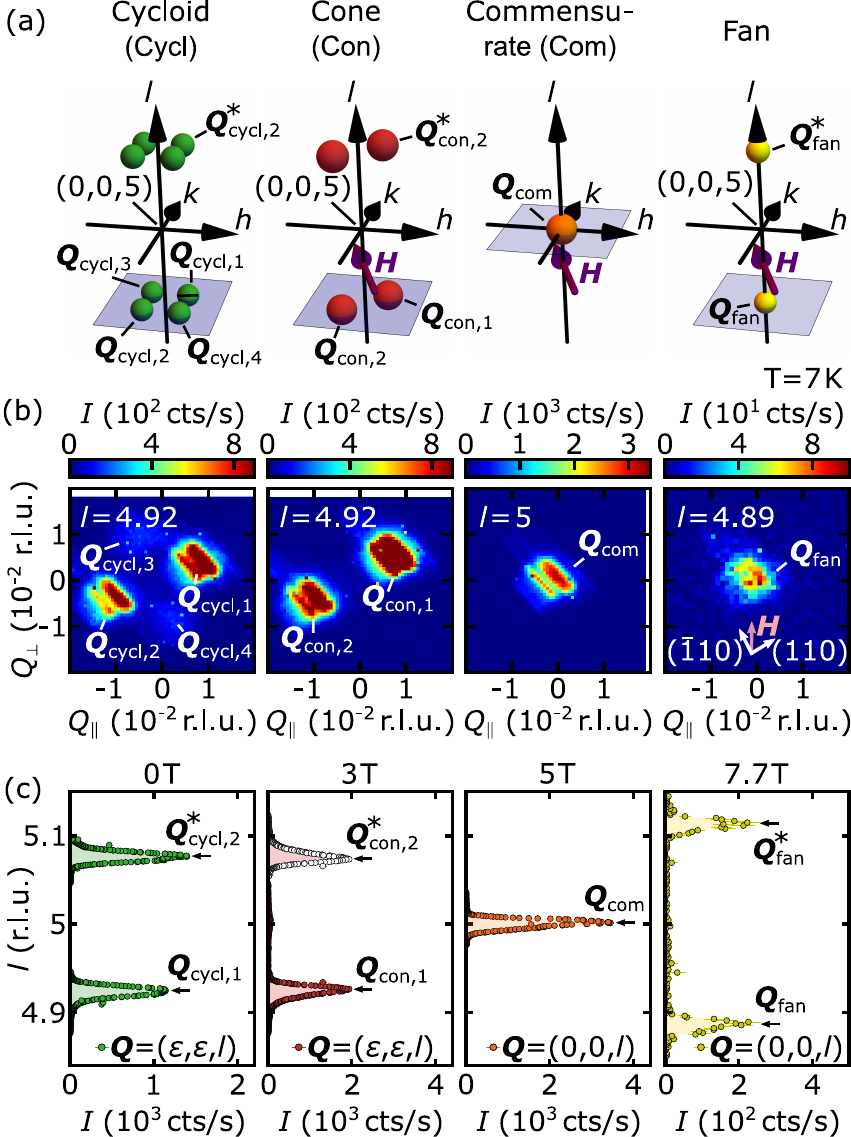}
\caption{Resonant elastic X-ray scattering. (a)~Schematic depiction of the REXS intensity around the reciprocal-space position $(h,k,l) = (0,0,5)$ in the four ordered phases established in bulk measurements~\cite{2022_Bauer_PhysRevMater}. Maxima at positions $\bm{Q}$ are indexed with the name of the phase. Arabic numbers indicate crystallographically equivalent positions attributed to different magnetic domains. Positions $\bm{Q}$ and $\bm{Q}^{*}$ are mirrored with respect to $(0,0,5)$ and belong to the same domain. Magnetic field $\bm{H}$ was applied along a nonsymmetry axis within the basal plane in order to discriminate single-domain and multi-domain states by means of their evolution under field, cf.\ supplementary information for data with field parallel to $\langle110\rangle$~\cite{SOI}. (b)~Intensity distributions recorded across planes of constant $l$, marked by blue shading in (a). (c)~Intensity when scanning $l$ through characteristic magnetic Bragg peaks at constant $h$ and $k$. For clarity, in the conical phase data were mirrored at $l = 5$ (open symbols).}
\label{figure2}
\end{figure}

The sample was cooled using a variable temperature insert. A cryomagnet was used to apply vertical magnetic fields of up to 14~T, where two field orientations were studied. In a first experiment, the magnetic field was applied in the tetragonal basal plane enclosing an angle of 20~deg with the $[\bar{1}10]$ axis. This way, the magnetic scattering of all domains could be studied in a single scattering channel, namely $\pi\rightarrow\sigma'$. In a second experiment, the magnetic field was applied parallel to the crystallographic $[\bar{1}10]$ axis, for which the evolution of domain populations as a function of field permitted to discriminate multi-$\bm{k}$ from single-$\bm{k}$ characteristics~\cite{SOI, 2022_Bauer_PhysRevMater}. The FLPA was carried out for this high-symmetry configuration. Data shown in Figs.~\ref{figure1} and \ref{figure2} were recorded with the first configuration; data shown in Fig.~\ref{figure3} were recorded with the second configuration; further data are shown in the supplementary information~\cite{SOI}. For clarity, momentum transfers are given in reciprocal lattice units (r.l.u.), corresponding to $\frac{2\pi}{a}$ along directions $h$ and $k$ or $\frac{2\pi}{c}$ along $l$.

For all four antiferromagnetic phases, REXS intensity was recorded at specific positions $\bm{Q}$ in the vicinity of the reciprocal-space position $(h,k,l) = (0,0,5)$, which is crystallographically forbidden. As shown in Fig.~\ref{figure1}(c), the integrated scattering intensities as a function of field reflect accurately the phase boundaries of the magnetic phase diagram. The intensity distributions are depicted schematically in Fig.~\ref{figure2}(a). Typical REXS data are presented in the form of two-dimensional maps inferred from scans at constant $l$ in Fig.~\ref{figure2}(b) and scans along $l$ at fixed $h$ and $k$ in Fig.~\ref{figure2}(c). The data presented below were measured under a rotation of the linear polarization by 90~deg, namely in the $\pi\rightarrow\sigma'$ channel, characteristic of magnetic scattering~\cite{1996_Hill_ActaCrystallogrA}.

In phase A ($H<H_1$), eight magnetic satellites were observed at $(\pm\epsilon,\pm\epsilon,5\pm\delta_{1})$ with $\epsilon=0.007(1)$ and $\delta_{1}=0.077(6)$ (green spheres). This field distribution implies superlattice modulations of the staggered magnetization with $c/(2\delta_{1})\approx 64~\mathrm{\AA}$ along $[001]$ and $a/(2\sqrt{2}\epsilon)\approx 215~\mathrm{\AA}$ in the basal plane along $\langle110\rangle$. As satellites antipodal to $(0,0,5)$ arise from the same domain of the incommensurate modulation axis, four crystallographically equivalent domains are distinguished that were populated equally after zero-field cooling. Maxima at reciprocal space positions with $l < 5$ are labeled by an index enumerating the domains. Maxima at $l > 5$ attributed to the same domain are denoted by an asterisk. In the $\pi\rightarrow\sigma'$ polarization channel, the maxima at $\bm{Q}_{\mathrm{cycl},3}$ and $\bm{Q}_{\mathrm{cycl},4}$ are weak due to well understood polarization effects for the scattering geometry chosen here, although all domains are populated equally (see supplementary information~\cite{SOI}).

\begin{figure*}
\centering	
\includegraphics{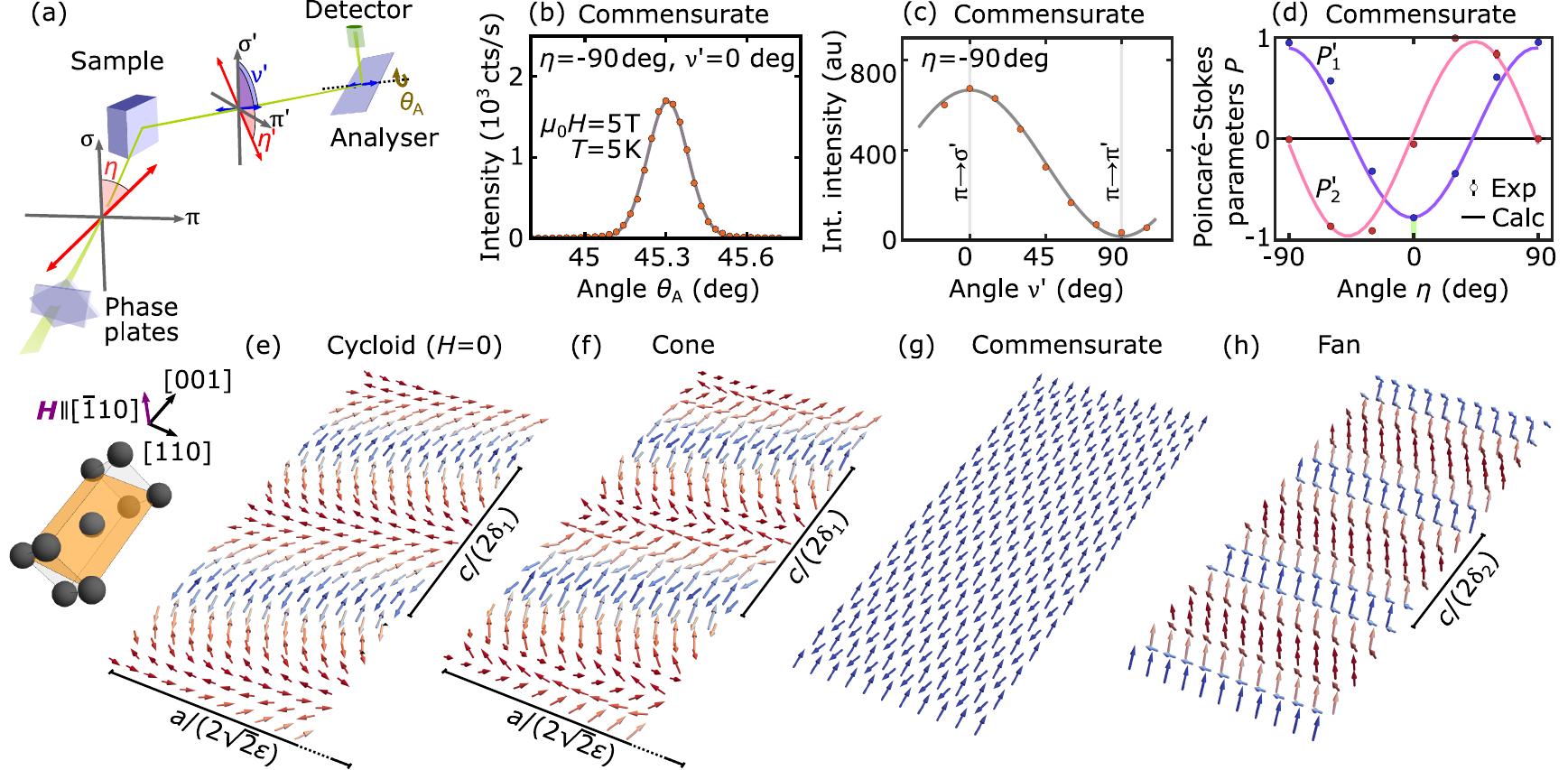}
\caption{Magnetic structure refinement by means of FLPA. (a)~Schematic depiction of the setup used for FLPA. Variables without and with prime denote quantities before and after scattering. Polarization directions $\pi$ and $\sigma$ are in and perpendicular to the scattering plane, respectively. The direction of the X-ray polarization (red double-headed arrow) with respect to $\sigma$ is denoted by the angle $\eta$. The rotatable analyzer crystal selects a polarization direction enclosing an angle $\nu'$ with $\sigma'$. The crystal may be rocked by the angle $\mathit{\Theta}_{A}'$. Phase plates determine the incident polarization~\cite{2013_Francoual_JPhysConfSer}. (b)~Rocking scan of the analyzer for a given incident ($\pi$) and scattered ($\sigma'$) polarization channel. Integrated intensity is inferred from a Gaussian fit (solid line). Typical data for the commensurate phase are shown. (c)~Integrated intensity for a given incident polarization ($\pi$) as a function of the analyzer orientation $\nu'$ and Poincare--Stokes fit (solid line). Magnetic intensity in channels $\pi\rightarrow\sigma'$ and $\pi\rightarrow\pi'$ reflect magnetization components in and perpendicular to the scattering plane. (d)~Poincare--Stokes parameters $P'_{1}$ and $P'_{2}$ as a function of the incident polarization angle $\eta$. Solid lines correspond to calculations based on commensurate antiferromagnetic order. Discrepancy from $P'_{1} = -1$ at $\eta=0$, marked in green, is attributed to charge scattering. (e)--(h)~Schematic real-space depictions of the magnetic structure in the different phases in the crystallographic $(1\bar{1}0)$ plane (orange shading). Blue and red colors indicate large and small components along $[001]$. The modulation length refers to the staggered magnetization.}
\label{figure3}
\end{figure*}

The scattering intensity in phase B ($H_1<H<H_2$) is characteristic of domains with an in-plane modulation perpendicular to the field (red spheres). In this field range, satellites at $\bm{Q}_{\mathrm{cycl},3}$ and $\bm{Q}_{\mathrm{cycl},4}$ vanish, while the modulation lengths remain unchanged $c/(2\delta_{1})\approx 64~\mathrm{\AA}$ along $[001]$ and $a/(2\sqrt{2}\epsilon)\approx 215~\mathrm{\AA}$ in the basal plane along $\left\langle110\right\rangle$. Accordingly, the in-plane modulation remains aligned with the crystallographic axes rather than following the low-symmetry field direction. The domain populations display hysteresis as a function of field, as illustrated in the supplementary information~\cite{SOI}.

Other than in phases A and B, scattering intensity in phase C ($H_2<H<H_3$) was only observed at $(0,0,5)$ (orange sphere), characteristic of single-domain commensurate antiferromagnetic order without superlattice modulations. Entering phase D ($H_3<H<H_4$), weak magnetic intensity at $(0,0,5\pm\delta_{2})$ with $\delta_{2} = 0.114(6)$ is observed (yellow spheres), characteristic of single-domain incommensurate order with a modulation length $43~\mathrm{\AA}$ of the staggered magnetization along $[001]$ and no superlattice modulation in the basal plane. For fields exceeding the highest critical fields observed in the bulk properties, i.e., $H_4<H$, no scattering intensity was observed as expected of the field-polarized state.

To determine the nature of the magnetic order unambiguously, FLPA was carried out in each magnetic phase for magnetic field parallel to $[\bar{1}10]$~\cite{2007_Mazzoli_PhysRevB, 2009_Hatton_JMagnMagnMater}. The experimental setup is schematically depicted in Fig.~\ref{figure3}(a). The procedure is illustrated by means of data recorded in the commensurate phase shown in Figs.~\ref{figure3}(b) to \ref{figure3}(d), cf.\ supplementary information for data recorded in the other phases~\cite{SOI}. For a given polarization angle $\eta$ of the incident beam, the scattering intensity for a given orientation of the analyzer crystal, $\nu'$, was determined by integrating over a rocking scan of the analyzer crystal using a Gaussian fit~\cite{2004_Detlefs_PhysicaB} [Fig.~\ref{figure3}(b)]. Such rocking scans were carried out for a series of analyzer orientations $\nu'$. Fitting the integrated intensities with the equation $f(\nu') \propto 1 + P'_{1}\cos2\nu' + P'_{2}\sin2\nu'$ [Fig.~\ref{figure3}(c)], the linear polarization of the scattered beam was determined in terms of its Poincare--Stokes parameters $P'_{1}$ and $P'_{2}$. This measurement protocol was repeated for different values of $\eta$~\cite{2012_Detlefs_EurPhysJSpecTop}. Finally, starting from the irreducible representations, values of the Poincar\'{e}--Stokes parameters $P'^{\mathrm{calc}}_1\left(\eta_i\right)$ and $P'^{\mathrm{calc}}_2\left(\eta_i\right)$ were calculated for each candidate magnetic structure and compared with the seven pairs of Poincar\'{e}--Stokes parameters experimentally determined [Fig.~\ref{figure3}(d)]. 

Crucial for the refinement of the magnetic structure, the FLPA allowed to single out the spin scattering contributions. Namely, in all magnetic phases scattering intensity was also observed under unchanged linear polarization, i.e., special care has to be taken to distinguish magnetic from nonmagnetic scattering contributions. In the $\sigma\rightarrow\sigma'$ channel, the scattering must be purely nonmagnetic, while it may be magnetic or nonmagnetic in the $\pi\rightarrow\pi'$ channel~\cite{1988_Blume_PhysRevB, 1996_Hill_ActaCrystallogrA}. For the magnetic structure refinement, it was assumed that the nonmagnetic scattering is due to charge scattering. For increasing magnetic field going from phase A to phase D, inclusion of the charge scattering improved the goodness of fit dramatically, cf.\ supplementary information~\cite{SOI}. In addition, intensity maxima were observed in the cycloidal and conical phase in the $\pi\rightarrow\pi'$ channel at $(0,0,5\pm\delta_{1})$, independent of the magnetic satellites. This intensity may be the characteristic of so-called truncation rods arising from finite penetration depth or a symmetry reduction due to structural modulations or charge-density wave order~\cite{2022_Moya_PhysRevMater}. While further studies are needed to resolve the origin of the charge scattering, determination of the magnetic structures pursued here turns out to be robust.

The magnetic structures inferred from REXS with FLPA, taking charge scattering into account, are depicted schematically for the crystallographic $(1\bar{1}0)$ plane in Figs.~\ref{figure3}(e) to \ref{figure3}(h). Starting with phase A, shown in Fig.~\ref{figure3}(e), type A antiferromagnetism is observed with an antiparallel coupling of the moments along the $\langle111\rangle$ directions and a long-wavelength cycloidal superstructure. The superstructure exhibits modulations along $[001]$ and one of the $\langle110\rangle$ axes. This superstructure supports four equivalent domain populations in zero field. Fig.~\ref{figure3}(e) depicts the domain associated with $\bm{Q}_{\mathrm{cycl,1}}$. Considering phase B, depicted in Fig.~\ref{figure3}(f), the same type A antiferromagnetism persists with a superstructure that is closely related to the cycloid, supporting modulations along $[001]$ and perpendicular to the field direction. The main difference with respect to phase A is the uniform magnetization along the field direction. Thus, with increasing field the opening angle of the conical structure decreases.

Phase C, shown in Fig.~\ref{figure3}(g), represents commensurate type A antiferromagnetism in which ferromagnetic layers of moments parallel and antiparallel to the $[001]$ axis alternate along the same axis, superimposed with a uniform magnetization along the field direction. The resulting magnetic structure is noncollinear but coplanar without additional twisting or scalar spin chiralities. Finally, as shown in Fig.~\ref{figure3}(h), phase D corresponds to type A order with a long-wavelength amplitude-modulated superstructure of moments pointing along $[001]$ and a uniform magnetization along the field direction. This modulation may be referred to as fan-like and differs distinctly from the cycloidal and conical modulations.

Considering the DM vectors permitted by the crystal structure~\cite{2022_Bauer_PhysRevMater}, Hamiltonian contributions of magnetic moments for the next-nearest neighbor bonds along $\langle111\rangle$ perpendicular to the field direction $[\bar{1}10]$ favor spin canting around $[\bar{1}10]$, such as in the cycloidal and conical state. For next-nearest neighbor bonds along $\langle111\rangle$ perpendicular to $[110]$, instead spin canting around $[110]$ is favored, as observed in the fan-like phase. In combination with the Zeeman energy in applied fields, modulated states as different as the cycloidal and the fan-like state may be realized.

We finally note that the critical field of the field-polarized state of order 9~T, which sets the scale of the antiferromagnetic exchange interactions, exhibits comparatively small anisotropy~\cite{2010_Kumar_PhysRevB, 2022_Bauer_PhysRevMater}. As the commensurate phase is encompassed by the phases supporting cycloidal, conical, and fan-like superstructures, the DM spin--orbit coupling must be comparable in strength to the antiferromagnetic exchange. Thus, building on the advantages offered by REXS with FLPA in studies of antiferromagnetic superstructures and materials not amenable to neutron scattering, we identify a highly unusual combination of interactions and magnetic phases which, to the best of our knowledge, has neither been reported experimentally nor addressed theoretically before.

\begin{acknowledgments}
We wish to thank M.\ Azhar, M.\ Garst, F.\ Haslbeck, J.\ R.\ Linares Mardegan, S.\ Mayr, A.\ Senyshyn, S.\ Sorn, and M.\ Wilde for fruitful discussions and assistance with the experiments. This study has been funded by the Deutsche Forschungsgemeinschaft (DFG, German Research Foundation) under TRR80 (From Electronic Correlations to Functionality, Project No.\ 107745057, Project E1), SPP2137 (Skyrmionics, Project No.\ 403191981, Grant PF393/19), and the excellence cluster MCQST under Germany's Excellence Strategy EXC-2111 (Project No.\ 390814868). Financial support by the European Research Council (ERC) through Advanced Grants No.\ 291079 (TOPFIT) and No.\ 788031 (ExQuiSid) as well as through the European Unions Horizon 2020 research and innovation program under the Marie Sk{\l}odowska--Curie grant agreement No.\ 884104 (PSI-FELLOW-III-3i) is gratefully acknowledged. We acknowledge DESY (Hamburg, Germany), a member of the Helmholtz Association HGF, for the provision of experimental facilities. Parts of this research were carried out at beamline P09 at PETRA III at DESY. Beamtimes were allocated for proposals I-20180440 and I-20200748.
\end{acknowledgments}

\end{document}


\title{Supplementary Material for \\
Resonant elastic x-ray scattering of antiferromagnetic superstructures in EuPtSi$_{\text{3}}$}

\author{Wolfgang~Simeth}
 \altaffiliation[Present address: ]{Paul Scherrer Institut (PSI), CH-5232 Villigen, Switzerland}
 \email{wolfgang.simeth@psi.ch}
 \affiliation{Physik-Department, Technische Universit\"at M\"unchen, D-85748 Garching, Germany}
 \affiliation{Laboratory for Neutron and Muon Instrumentation, Paul Scherrer Institute, Villigen PSI, Switzerland}

\author{Andreas~Bauer}
 \affiliation{Physik-Department, Technische Universit\"at M\"unchen, D-85748 Garching, Germany}
 \affiliation{Zentrum f\"ur QuantumEngineering (ZQE), Technische Universit\"at M\"unchen, D-85748 Garching, Germany}

\author{Christian~Franz}
 \affiliation{Physik-Department, Technische Universit\"at M\"unchen, D-85748 Garching, Germany}
 \affiliation{J\"ulich Centre for Neutron Science (JCNS) at Heinz Maier-Leibnitz Zentrum (MLZ), D-85748 Garching, Germany}

\author{Aisha~Aqeel}
 \affiliation{Physik-Department, Technische Universit\"at M\"unchen, D-85748 Garching, Germany}

\author{Pablo~J.~Bereciartua Perez}
 \affiliation{Deutsches Elektronen-Synchrotron (DESY), D-22607 Hamburg, Germany}

\author{Jennifer~A.~Sears}
 \affiliation{Deutsches Elektronen-Synchrotron (DESY), D-22607 Hamburg, Germany}

\author{Sonia~Francoual}
 \affiliation{Deutsches Elektronen-Synchrotron (DESY), D-22607 Hamburg, Germany}

\author{Christian~H.~Back}
 \affiliation{Physik-Department, Technische Universit\"at M\"unchen, D-85748 Garching, Germany}
 \affiliation{Munich Center for Quantum Science and Technology (MCQST), Technische Universit\"at M\"unchen, D-85748 Garching, Germany}

\author{Christian~Pfleiderer}
 \affiliation{Physik-Department, Technische Universit\"at M\"unchen, D-85748 Garching, Germany}
 \affiliation{Zentrum f\"ur QuantumEngineering (ZQE), Technische Universit\"at M\"unchen, D-85748 Garching, Germany}
 \affiliation{Munich Center for Quantum Science and Technology (MCQST), Technische Universit\"at M\"unchen, D-85748 Garching, Germany}

\begin{abstract}
This Supplementary Material contains additional information on the experimental and computational methods. After a short account on the single-crystal samples, details on the resonant elastic X-ray scattering experiment are presented including information on the mathematical framework employed for the description of the magnetic scattering contributions. Following a short summary of the fundamentals of A-type antiferromagnetism in a tetragonal $I4mm$ crystal structure, complementary scattering data for magnetic field applied along the $[\bar{1}10]$ axis are presented. The propagation vectors and their assignment to different magnetic domains are given in tabular form and the dependence of the magnetic order on the temperature and field history is discussed. Subsequently, after starting with an analysis of the direct beam as a point of reference, the full linear polarization analysis of the four long-range ordered phases and the determination of magnetic structures is elaborated on. 
\end{abstract}

\maketitle

\clearpage

\section{Sample preparation}

The resonant elastic X-ray scattering (REXS) experiments in our study were carried out on a single-crystal cube with an edge length of 2~mm that was prepared from a larger, high-quality single-crystal ingot grown by means of the optical floating-zone technique~\cite{2011_Neubauer_RevSciInstrum, 2016_Bauer_RevSciInstruma, 2022_Bauer_PhysRevMaterials}. The cube possessed surfaces perpendicular to $[001]$, $[110]$, and $[\bar{1}10]$, respectively. For the REXS measurements, one of the surfaces perpendicular to $[001]$ was polished mechanically. The magnetic phase diagram, shown in Fig.~1(c) of the main text, 
was also measured on this cube, cf.\ Ref.~\cite{2022_Bauer_PhysRevMaterials}. Room-temperature X-ray diffraction of a powder prepared of the same float-zoned ingot used to obtain the single crystal examined in our study was reported in Ref.~\cite{2022_Bauer_PhysRevMaterials}, indicating high sample quality. Neutron powder diffraction, also reported in Ref.~\cite{2022_Bauer_PhysRevMaterials}, as measured between 4\,K and 300\,K did not provide any evidence suggestive of a structural transition.

\clearpage

\section{Resonant elastic X-ray scattering}

The REXS experiments were carried out in the second experimental hutch of the resonant scattering and diffraction beamline P09 at the synchrotron source PETRA~III~\cite{2013_Strempfer_JSynchrotronRad}. This hutch is equipped with a heavy-load diffractometer. A split-pair vertical field cryomagnet with a variable temperature insert permits us to study the entire magnetic phase diagram of EuPtSi$_{3}$. Magnetic structures were determined by means of a full linear polarization analysis using a double phase-retarder in combination with an appropriate analyzer~\cite{2013_Francoual_JPhysConfSer}.

At the L$_{\mathrm{II}}$ resonance (7.61~keV) of europium, on which REXS was performed, the (006) reflection of the pyrolithic graphite (PG) analyzer crystal used for our experiments appears under the scattering angle $2\theta_{\mathrm{A}}=90.34\,\mathrm{deg}$. Analyzer leakage of transmitted finite polarization contributions in the scattering plane of the analyzer crystal is expected to be smaller than $\cos^2\left(2\theta_{\mathrm{A}}\right)=0.004\,\%$.

Two sets of REXS experiments were performed, namely (i)~with the magnetic field parallel to a nonsymmetry in-plane direction enclosing an angle of $20^{\circ}$ with the $[\bar{1}10]$ axis and (ii)~with the vertical magnetic field parallel to the $[\bar{1}10]$ axis of the sample. For the second configuration, as presented in Fig.~\ref{figureS2}, the sample was mounted on a copper holder such that $\left[\bar{1}10\right]$ was vertical and $\left[110\right]$ as well as $\left[001\right]$ were in the horizontal scattering plane of the diffractometer. Scattering was performed from the polished $\left(001\right)$ surface of the cube-shaped sample. The orientation matrix was refined on the structural peaks $\left(0,0,6\right)$ and $\left(-1,-1,8\right)$, yielding tetragonal lattice constants  $a=4.26\,\mathrm{\AA}$ and $c=9.79\,\mathrm{\AA}$, in good agreement with previous reports~\cite{2010_Kumar_PhysRevB}. For the first configuration, as presented in Fig.~1 of the main text, the cubic sample was mounted on a copper holder such that the $\left[001\right]$ direction and a direction in the $ab$-plane that encloses an angle of $\approx20$ deg with $\left[\bar{1}10\right]$ were in the horizontal scattering plane. Scattering was performed again from the polished $\left(001\right)$ surface of the cube-shaped sample.

\clearpage

\section{Polarized resonant elastic X-ray scattering}

In the following, a framework is presented that allows for the mathematical description of polarized resonant elastic X-ray scattering. The Poincar\'{e}--Stokes parameters $P_1$, $P_2$, and $P_3$, summarized as the vector $\boldsymbol{P}=\left(P_1,P_2,P_3\right)$, provide a full characterization of the polarization state of a photon \cite{2012_Detlefs_EurPhysJSpecTop}. The parameters $P_1$ and $P_2$ describe the state of linear polarization, where the degree of linear polarization is given by $\sqrt{P_1^2+P_2^2}$. The parameter $P_3$ describes the state of circular polarization. The linearly polarized X-ray beam in our study displays almost 100 \% polarization, as discussed in Sec.~\ref{section:FLPAdirectbeam}. In this case of ideal polarization, the Poincar\'{e}--Stokes parameters are given by:
\begin{align}
	\boldsymbol{P}=\left(P_1,P_2,P_3\right)=\left(\cos\left(2\eta\right),\sin\left(2\eta\right),0\right) \,   . \label{equation:PoincareStokesIncidentIdeal}
\end{align}

For our calculation, it is convenient to treat the X-ray beam polarization in the framework of the density-matrix (or coherency-matrix) formalism~\cite{1988_Blume_PhysRevB, 2012_Detlefs_EurPhysJSpecTop}, in which a X-ray beam with Poincar\'{e}--Stokes vector $\boldsymbol{P}$ may be represented by the matrix:
\begin{align}
	\mu=\frac{1}{2}\left[ \sigma_0 +  \boldsymbol{\sigma}\cdot \boldsymbol{P}\right]=\frac{1}{2}\cdot\begin{pmatrix}
		1+P_1  &   P_2-\mathrm{i}P_3   \\
		P_2+\mathrm{i}P_3   &  1-P_1
	\end{pmatrix}   \, .
\end{align}
Here, $\sigma_0$ denotes the $2\times2$ identity matrix and the vector $\boldsymbol{\sigma}$ contains the Pauli matrices in the form
\begin{align}
	\sigma_1=\begin{pmatrix} 1 & 0\\ 0 & -1 \end{pmatrix}, \, \,  \sigma_2=\begin{pmatrix} 0 & 1\\ 1 & 0 \end{pmatrix} , \, \,  \sigma_3=\begin{pmatrix} 0 & -\mathrm{i}\\ \mathrm{i} & 0 \end{pmatrix}   \, .
	\nonumber
\end{align}

\begin{figure}[h]
	\includegraphics{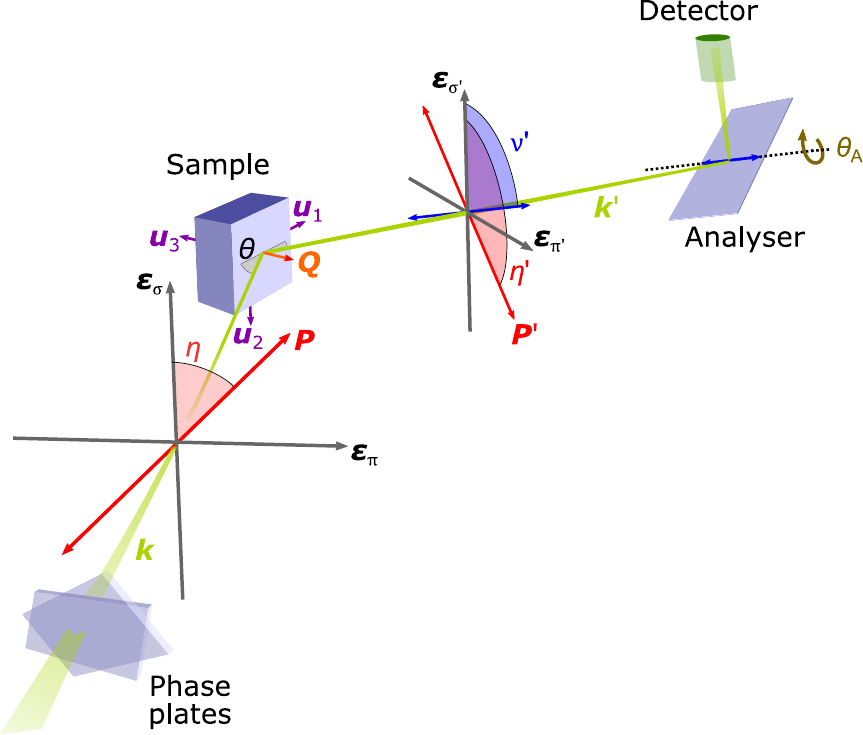}
	\caption{\label{figureS1}Schematic depiction of the experimental setup and the scattering geometry, corresponding to a more complex version of Fig.~3(a) in the main text. The vectors $\boldsymbol{u}_1$, $\boldsymbol{u}_2$, $\boldsymbol{u}_3$, $\boldsymbol{\epsilon}_{\sigma}$, $\boldsymbol{\epsilon}_{\pi}$, $\boldsymbol{\epsilon}_{\sigma'}$, and $\boldsymbol{\epsilon}_{\sigma'}$ are presented relative to the directions $\boldsymbol{k}$ and $\boldsymbol{k}'$ of incident and scattered X-rays. The linear polarizations of incident and scattered X-rays with Poincar\'{e}--Stokes vectors $\boldsymbol{P}$ and $\boldsymbol{P}'$ are shown as red double-headed arrows. The axis of linear polarization that is transmitted by the analyses is indicated by a blue double-headed arrow. See text for details.}
\end{figure}

For the evaluation of scattering amplitudes in our study, the reference frame as introduced by Blume and Gibbs~\cite{1988_Blume_PhysRevB} was used that is given by
\begin{align}
	\boldsymbol{u}_1&= \left(\hat{\boldsymbol{k}}+\hat{\boldsymbol{k}}'\right) / \left(2 \cos\theta \right)\\
	\boldsymbol{u}_2&= \left(\hat{\boldsymbol{k}}\times\hat{\boldsymbol{k}}'\right) /\, \left( 2 \sin2\theta \right) \\
	\boldsymbol{u}_3&= \left(\hat{\boldsymbol{k}}-\hat{\boldsymbol{k}}'\right) / \left(2 \sin\theta \right) \, .
\end{align}
Here, $\boldsymbol{k}$ and $\boldsymbol{k}'$ denote the wave vectors of incident and scattered X-rays, respectively, and the scattering angle $2\theta$ is defined by $\hat{\boldsymbol{k}}\cdot\hat{\boldsymbol{k}}'=\cos\left(2\theta\right)$. This geometry is illustrated in Fig.~\ref{figureS1}. The unit vector $\boldsymbol{u}_1$ lies in the scattering plane and perpendicular to the momentum transfer $\boldsymbol{Q}=\boldsymbol{k}'-\boldsymbol{k}$. The unit vector $\boldsymbol{u}_2$ is perpendicular to the scattering plane. The unit vector $\boldsymbol{u}_3$ lies in the scattering plane being antiparallel to $\boldsymbol{Q}$. 

In turn, the vectors $\boldsymbol{\epsilon}_{\sigma}=-\boldsymbol{u}_2$, $\boldsymbol{\epsilon}_{\pi}=\sin(\theta)\boldsymbol{u}_1-\cos(\theta)\boldsymbol{u}_3$, $\boldsymbol{\epsilon}_{\sigma'}=-\boldsymbol{u}_2$, and $\boldsymbol{\epsilon}_{\pi'}=-\sin(\theta)\hat{\boldsymbol{u}}_1-\cos(\theta)\boldsymbol{u}_3$ define the axes $\sigma$, $\pi$, $\sigma'$, and $\pi'$, respectively. Hence, the X-ray scattering amplitudes may be written using a two-component matrix formalism, in which the entries are associated with the polarization transition channels as follows~\cite{2012_Detlefs_EurPhysJSpecTop}
\begin{align}
	\begin{pmatrix}
		\sigma\rightarrow \sigma'  &  \pi\rightarrow \sigma'  \\
		\sigma\rightarrow \pi'   &  \pi\rightarrow \pi'
	\end{pmatrix}   \, .
\end{align}

When the Fourier transform of the scattering amplitude for a given scattering process is represented by the $2\times2$ matrix $M\left(\boldsymbol{Q}\right)$, the density matrix $\mu'$, the intensity $I\left(\boldsymbol{Q}\right)$, and the Poincar\'{e}--Stokes vector $\boldsymbol{P}'$ of the scattered beam are given by~\cite{1988_Blume_PhysRevB, 1957_Fano_RevModPhys, 2008_vanderLaan_ComptesRendusPhysique}
\begin{align}
	\mu'&=M\left(\boldsymbol{Q}\right)\mu M^{\dagger}\left(\boldsymbol{Q}\right) \, , \nonumber\\
	I\left(\boldsymbol{Q}\right) &=\mathrm{Tr}\left(\mu'\right) \, , \nonumber \\
	\boldsymbol{P}' &=    \frac{\mathrm{Tr}\left(\boldsymbol{\sigma}\cdot \mu'\right)} {\mathrm{Tr}\left(\mu'\right)} \, .
	\label{equation:poincarestokesofscatteredbeam}
\end{align}

The magnetic cross section for resonant X-rays is typically dominated by electric dipole transitions (E1), cf.\ for instance Ref.~\cite{1996_Hill_ActaCrystA}. For magnetically ordered ions in the magnetic structure $\boldsymbol{m}\left(\boldsymbol{R}\right)$, the resonant dipole scattering amplitude may be written as~\cite{1996_Hill_ActaCrystA}
\begin{align}
	f_{nE1}&=    \rho F^{\left(0\right)}\cdot \begin{pmatrix}
		1  &  0  \\
		0   &  \cos\left(2\theta\right)
	\end{pmatrix}  - \nonumber  \\
	&-\mathrm{i}\cdot F^{\left(1\right)}\cdot \begin{pmatrix}
		0  &  z_1\cdot\cos\left(\theta\right) + z_3\cdot\sin\left(\theta\right) \nonumber \\
		z_3\cdot\sin\left(\theta\right)-z_1\cdot\cos\left(\theta\right)   & -z_2 \sin\left(2\theta\right)
	\end{pmatrix} + \nonumber\\
	&+ F^{\left(2\right)}\cdot \begin{pmatrix}
		z_2^2  &  -z_2\cdot \left( z_1\cdot\sin\left(\theta\right) - z_3\cdot\cos\left(\theta\right)\right) \\
		z_2\cdot\left(z_1\cdot\sin\left(\theta\right)+z_3\cdot\cos\left(\theta\right)\right)   & -\cos^2\left(\theta\right) \left(z_1^2\tan^2\left(\theta\right)+z_3^2\right)
	\end{pmatrix}   \,    ,
	\label{equation:resonantdipolescatteringamplitude}
\end{align}
with $ \rho$ denoting the charge density and the vector $\boldsymbol{m}^{\boldsymbol{u}}\left(\bm{R}\right)=\left(z_1,z_2,z_3\right)$ denoting the magnetization vector in the basis $\boldsymbol{u}_1$, $\boldsymbol{u}_2$, $\boldsymbol{u}_3$.


The first term is associated with charge scattering, whereas the second and third terms describe magnetic scattering. In an incommensurate antiferromagnet with wave vector $\boldsymbol{\tau}$, the second term typically produces scattering intensity at first-order peak positions $\boldsymbol{Q}=\boldsymbol{G}\pm \boldsymbol{\tau}$ and the third term at second-order peak positions $\boldsymbol{Q}=\boldsymbol{G}\pm 2\cdot\boldsymbol{\tau}$ around structural peaks $\boldsymbol{G}$. In addition, when the magnetic structure has a contribution with wave-vector $\bm{k}=0$, which in EuPtSi$_3$ is the case at finite magnetic field, the third term may produce a mixed contribution that emerges from the presence of both ferromagnetic and antiferromagnetic wave-vectors and that leads to magnetic intensity at first-order peak positions $\boldsymbol{Q}=\boldsymbol{G}\pm \boldsymbol{\tau}$~\cite{1996_Hill_ActaCrystA,1994_Pengra_JPhysCondensMatter}. The charge-scattering term also can lead to intensity at first order magnetic Bragg peak positions, when the Fourier transform $\rho(\bm{Q})$ of the charge density is finite at $\bm{Q}=\boldsymbol{G}\pm \boldsymbol{\tau}$, which may be the case in the presence of a charge-density wave or other forms of structural symmetry lowering.


\clearpage

\section{A-type antiferromagnetism in the space group $I4mm$}

As pointed out in the manuscript, magnetic structures in EuPtSi$_{3}$ may be described as variations of A-type antiferromagnetism. In the following, we present a brief account on A-type antiferromagnetism in the $I4mm$ space group and its characteristic signatures in diffraction experiments. For an atom X that occupies the Wyckoff position $a$ in the space group $I4mm$, such as Eu in EuPtSi$_{3}$, A-type antiferromagnetism may be described by $\left(001\right)$ layers that are stacked along the $\left[001\right]$ axis with a stacking distance of $c/2$, ferromagnetic intra-layer coupling, and antiferromagnetic inter-layer coupling. Alternatively, the three-dimensional body-centered tetragonal lattice formed by the atoms X may be described in terms of two primitive tetragonal sublattices connected by $\left\langle111\right\rangle$ bonds that feature ferromagnetic intra-sublattice coupling and antiferromagnetic inter-sublattice coupling. In diffraction experiments, A-type antiferromagnetism in the $I4mm$ space group is indicated by Bragg peaks at the M point in reciprocal space, which denotes the crystallographically forbidden position $\left(1,1,1\right)$ in reciprocal space. The point M is also equivalent to the positions $\left(0,0,u\right)$ when $u$ is an odd integer.

\clearpage

\section{Complementary diffraction data for field along $\left[\bar{1}10\right]$}

Magnetic order in EuPtSi$_{3}$ for field perpendicular to the $\left[001\right]$ axis was investigated in two experimental configurations. In the set of experiments that are presented and discussed in Figs.~1 and 2 of the main text, the magnetic field was applied along an axis in the basal plane that encloses an angle of $\approx 20\,\mathrm{deg}$ with the $\left[\bar{1}10\right]$ direction. As major advantage of this configuration, all magnetic domains can be studied in the $\pi\rightarrow\sigma'$ channel, in which all scattering intensity was magnetic. In the set of experiments that are presented in the following, the magnetic field was applied along the $\left[\bar{1}10\right]$ direction. This configuration is ideally suited for studying the domain populations in the cycloidal and conical phases as in the case of EuPtSi$_{3}$ the symmetry breaking between the different domains for field perpendicular to $\left[001\right]$ is most pronounced for field along a $\left\langle110\right\rangle$ axis. Unless stated else, data were recorded after initial zero-field cooling and subsequently applying the field values stated.  

\begin{figure}
	\includegraphics{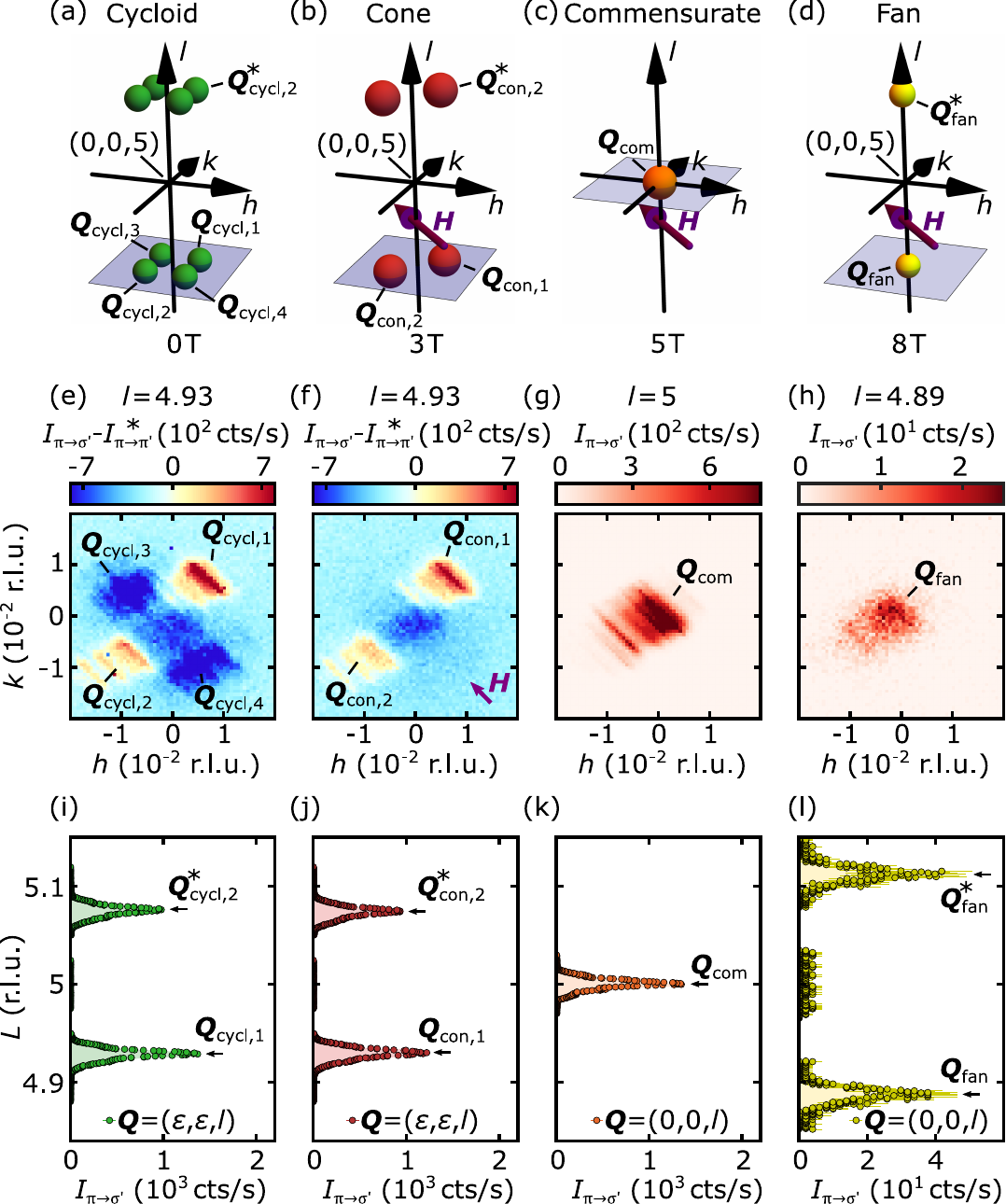}
	\caption{\label{figureS2}Resonant elastic X-ray scattering from magnetic order in EuPtSi$_{3}$ for field along $\left[\bar{1}10\right]$. \mbox{(a)--(d)}~Schematic depiction of the REXS intensity in the four ordered phases in vicinity of the reciprocal space position $(h,k,l) = (0,0,5)$. Antiferromagnetic phases with cycloidal (green), conical (red), commensurate (orange), and fan-like (yellow) order are distinguished. \mbox{(e)--(h)}~Intensity distributions recorded across planes of constant $l$, marked by blue shading in (e)--(h). Data were measured by means of $h$-scans at constant $k$. In order to visualize the full scattering pattern, the colormaps for the cycloidal and conical phases correspond to the difference between the $\pi\rightarrow\sigma'$ and the $\pi\rightarrow\pi'$ channel (multiplied by a factor of 11.3). \mbox{(i)--(l)}~Intensity when scanning $l$ through characteristic magnetic Bragg peaks at constant $h$ and $k$. Data shown in (l) were recorded under decreasing magnetic field.}
\end{figure}

For both field directions studied, namely for field along an in-plane direction enclosing an angle of $\approx 20\,\mathrm{deg}$ with $\left[\bar{1}10\right]$ shown in Figs.~1 and 2 of the main text and for field along $\left[\bar{1}10\right]$ shown in Fig.~\ref{figureS2}, the scattering data and the magnetic phase diagram are highly reminiscent of each other. In analogy to Fig.~2(a) of the main text, a schematic depiction of the magnetic Bragg peaks in the vicinity of the reciprocal space position $\left(0,0,5\right)$ is illustrated for the four  ordered phases in Figs.~\ref{figureS2}(a) to \ref{figureS2}(d). In analogy to Figs.~2(b) and 2(c) of the main text, typical REXS data at constant $l$ and for a scan along $l$ are presented in Figs.~\ref{figureS2}(e) to \ref{figureS2}(h) and Figs.~\ref{figureS2}(i) to \ref{figureS2}(l), respectively. The color plots showing data at constant $l$ were constructed from REXS-data measured on discrete positions on a $(H,K)$ grid with a lattice spacing reflected by the pixel size in the color plots. Reciprocal space positions of the maxima presented in Fig.~\ref{figureS2} and Fig.~2 of the main text agree with each other within the error margins.

In the cycloidal phase, at zero magnetic field all magnetic domains were populated equally. The domains $\boldsymbol{Q}_{\mathrm{cycl},1}$ and $\boldsymbol{Q}_{\mathrm{cycl},2}$ possess modulations in the basal plane that are perpendicular to the field. These domains exhibit diffraction in the $\pi\rightarrow\sigma'$ channel, while their intensity is vanishingly small in the $\pi\rightarrow\pi'$ channel. For both the incident linear polarizations $\sigma$ and $\pi$, essentially the entire scattered polarization is linearly polarized. In turn, the domains at $\boldsymbol{Q}_{\mathrm{cycl},3}$ and $\boldsymbol{Q}_{\mathrm{cycl},4}$ with modulations parallel to the field possess substantial intensity in the $\pi\rightarrow\pi'$ channel and vanishingly small intensity in the $\pi\rightarrow\sigma'$ channel. For an incident polarization of $\pi$ the scattered beam exhibits essentially circular X-ray polarization. The maximum scattering intensity in the polarization channels $\pi\rightarrow\sigma'$ and $\pi\rightarrow\pi'$ observed for the latter two domains is distinctly smaller than for the domains with $\boldsymbol{Q}_{\mathrm{cycl},1}$ and $\boldsymbol{Q}_{\mathrm{cycl},2}$.

\begin{figure}
	\includegraphics{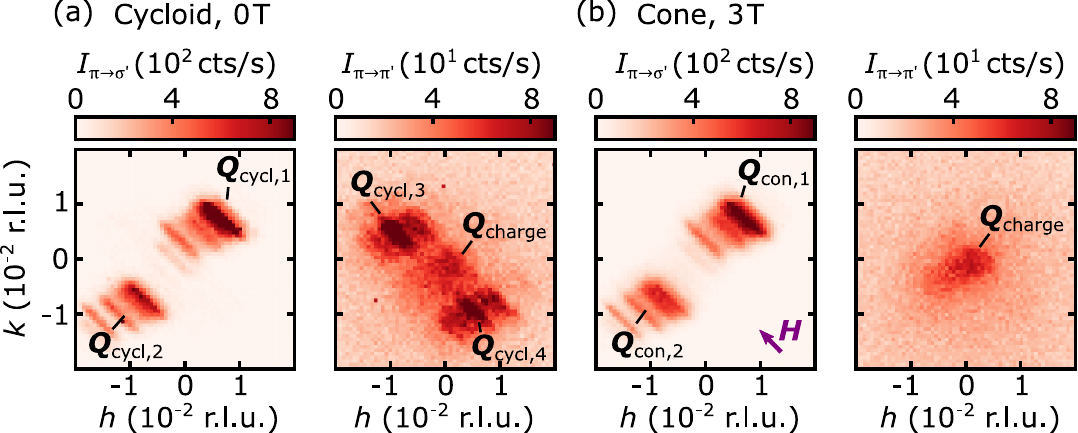}
	\caption{\label{figureS3}Comparison of the resonant elastic X-ray scattering intensity distributions between the $\pi\rightarrow\sigma'$ and $\pi\rightarrow\pi'$ channels for the cycloidal and the conical phase. (a) Resonant elastic X-ray scattering data recorded in the cycloidal phase at $l=4.93$ in the $\pi\rightarrow\sigma'$ channel (left) and in the $\pi\rightarrow\pi'$ channel (right). The intensity that appears at the position $\bm{Q}_{\mathrm{charge}} = (0,0,4.93)$ is attributed to charge scattering. (b) Resonant elastic X-ray scattering data recorded in the conical phase at $l=4.93$ in the $\pi\rightarrow\sigma'$ channel (left) and in the $\pi\rightarrow\pi'$ channel (right).}
\end{figure}

In order to visualize these aspects, the difference between the intensities in the $\pi\rightarrow\sigma'$ and the $\pi\rightarrow\pi'$ channel, normalized such that they display equal maxima, is shown in Fig.~\ref{figureS2}. Data corresponding to the separate channels are presented in Fig.~\ref{figureS3}.

In the conical phase, only the two domains at $\boldsymbol{Q}_{\mathrm{con},1}$ and $\boldsymbol{Q}_{\mathrm{con},2}$ with modulations in the basal plane that are perpendicular to the field were populated and featured intensity in the $\pi\rightarrow\sigma'$ channel. In contrast, the domains at $\boldsymbol{Q}_{\mathrm{cycl},3}$ and $\boldsymbol{Q}_{\mathrm{cycl},4}$ were depopulated as no REXS intensity was observed.

In the commensurate and the fan-like phases, magnetic structures with a single orientation of the wave vector are established, see main text for further details.

In the context of these supplementary data, we note that in the cycloidal and conical phase substantial scattering intensity was observed at the peak position $\bm{Q}_{\mathrm{charge}} = (0,0,4.93)$, as shown in the center of Figs.~\ref{figureS2}(e) and \ref{figureS2}(f) as well as in Fig.~\ref{figureS3}. This intensity does not appear in the $\pi\rightarrow\sigma'$ channel. This observation excludes a magnetic origin, as finite scattering intensity would be expected also in the $\pi\rightarrow\sigma'$ channel at zero magnetic field for any irreducible representation and equal distribution of magnetic domains. Consequently, this intensity may be attributed most likely to charge scattering and the lowering of structural symmetries as discussed below. In addition, we note that in the cycloidal and conical phases a uniformly distributed intensity of around 17(3) cts/s appears in the $\pi\rightarrow\pi'$ channel, which may be attributed to diffuse charge scattering~\cite{2018_Rahn_PhysRevB}.




\clearpage

\section{Propagation vectors and indexation of magnetic domains}

In the cycloidal, conical, commensurate, and fan-like phases, the antiferromagnetic propagation vectors in the conventional tetragonal basis may be described as $\boldsymbol{k}_{\mathrm{cycl}}=\left(\epsilon,\epsilon,1-\delta_{1}\right)$, $\boldsymbol{k}_{\mathrm{con}}=\left(\epsilon,\epsilon,1-\delta_{1}\right)$, $\boldsymbol{k}_{\mathrm{com}}=\left(0,0,1\right)$, and $\boldsymbol{k}_{\mathrm{fan}}=\left(0,0,1-\delta_{2}\right)$, respectively. In the following, the different possible orientations of these magnetic wave-vectors in tetragonal crystal symmetry and the resulting orientational magnetic domains are discussed.

The stars of the wave vectors in the cycloidal, conical, commensurate, and fan-like phases contain 4, 4, 1, and 1 propagation vectors, respectively, as summarized in Tab.~\ref{tableS1}. The wave vectors in the table refer to the conventional tetragonal basis. Different propagation vectors of the same star are associated with different orientational domains of a wave vector and hence of a magnetic structure. In contrast to the cycloidal and conical phases, where the crystal symmetries give rise to four orientational wave vector domains, the commensurate and fan-like phases exhibit only one orientational domain. 

For the four long-range ordered magnetic phases, the magnetic domains that were populated in the experiment with field parallel to $\left[\bar{1}10\right]$ are presented in Tab.~\ref{tableS2}. This table lists all magnetic peaks in the vicinity of the reciprocal space position $\left(0,0,5\right)$ that belong to one of the domains with finite scattering intensity, as also shown in Fig.~\ref{figureS2}. Peaks at incommensurate positions that are antipodal with respect to $\left(0,0,5\right)$ are associated with the same domain. 

\begin{table}
	\caption{\label{tableS1}Antiferromagnetic propagation vectors in the cycloidal (cycl), conical (con), commensurate (c), and fan-like (fan) phases. For each phase, the propagation vector star is given in the second row. The specific propagation vectors of each star are listed in rows 3 to 6. The incommensurate parameters are given by $\epsilon=0.007$, $\delta_1=0.073\approx \delta_{\mathrm{cycl}}\approx \delta_{\mathrm{con}}$, and $\delta_2=0.113= \delta_{\mathrm{fan}}$, cf.\ Tab.~\ref{tableS2}. All vectors in the table are given in the conventional tetragonal basis.}
	\begin{ruledtabular}
		\begin{tabular}{c|cccc}
			Phase     & Cycloid   & Cone & Commensurate & Fan   \\ \hline
			$\boldsymbol{k}$ star &  $ \left\lbrace \left(\epsilon,\epsilon,1-\delta_{1}\right)\right\rbrace$  & $ \left\lbrace \left(\epsilon,\epsilon,1-\delta_{1}\right)\right\rbrace$  & $ \left\lbrace \left(0,0,1\right)\right\rbrace$  & $ \left\lbrace \left(\epsilon,\epsilon,1-\delta_{2}\right)\right\rbrace$  \\   \hline
			$\boldsymbol{k}$ vectors    &    $ \boldsymbol{k}_{\mathrm{cycl,1}}=\left(\epsilon,\epsilon,1-\delta_{1}\right)$   & $ \boldsymbol{k}_{\mathrm{con,1}}=\left(\epsilon,\epsilon,1-\delta_{1}\right)$ & $ \boldsymbol{k}_{\mathrm{com}}=\left(0,0,1\right)$ &  $ \boldsymbol{k}_{\mathrm{fan}}=\left(\epsilon,\epsilon,1-\delta_{2}\right)$  \\
			&      $\boldsymbol{k}_{\mathrm{cycl,2}}=\left(-\epsilon,-\epsilon,1-\delta_{1}\right)$&    $\boldsymbol{k}_{\mathrm{con,2}}=\left(-\epsilon,-\epsilon,1-\delta_{2}\right)$ &      \\
			&      $\boldsymbol{k}_{\mathrm{cycl,3}}=\left(-\epsilon,\epsilon,1-\delta_{1}\right)$ &  $\boldsymbol{k}_{\mathrm{con,3}}=\left(-\epsilon,\epsilon,1-\delta_{1}\right)$&     \\
			&      $\boldsymbol{k}_{\mathrm{cycl,4}}=\left(\epsilon,-\epsilon,1-\delta_{1}\right)$ &  $\boldsymbol{k}_{\mathrm{con,4}}=\left(\epsilon,-\epsilon,1-\delta_{1}\right)$ &     
		\end{tabular}
	\end{ruledtabular}
\end{table}

\begin{table}
	\caption{\label{tableS2}Overview of the magnetic domains in the cycloidal (cycl), conical (con), commensurate (c), and fan-like (fan) phases that were populated in the experiment with field parallel to $\left[\bar{1}10\right]$. The list includes all Bragg peaks in the vicinity of the reciprocal space position $\left(0,0,5\right)$ that belong to one of the magnetic domains as shown in Fig.~\ref{figureS2}. For each wave vector $\boldsymbol{Q}$, the position in reciprocal space and the indexation are presented in the second and third column, respectively. The incommensurate parameters $\epsilon=0.007(1)$, $\delta_{\mathrm{cycl}}=0.073(5)$, $\delta_{\mathrm{con}}=0.072(5)$, and $\delta_{\mathrm{fan}}=0.113(7)$ were inferred from the location of Gaussian profiles fitted to the REXS data. The error corresponds to one standard deviation $\sigma$ of the fitted profiles.}
	\begin{ruledtabular}
		\begin{tabular}{ccc}
			$\boldsymbol{Q}$-label & reciprocal space coordinates &    indexation by $\boldsymbol{k}$ vector \\  \hline
			$\boldsymbol{Q}_{\mathrm{cycl},1}$ & $\left(\epsilon,\epsilon,5-\delta_{\mathrm{cycl}}\right)$ & $\left(0,0,4\right)+\boldsymbol{k}_{\mathrm{cycl},1}$  \\
			$\boldsymbol{Q}_{\mathrm{cycl},2}$ & $\left(-\epsilon,-\epsilon,5-\delta_{\mathrm{cycl}}\right)$ & $\left(0,0,4\right)+\boldsymbol{k}_{\mathrm{cycl},2}$   \\
			$\boldsymbol{Q}_{\mathrm{cycl},3}$ & $\left(-\epsilon,\epsilon,5-\delta_{\mathrm{cycl}}\right)$ & $\left(0,0,4\right)+\boldsymbol{k}_{\mathrm{cycl},3}$  \\
			$\boldsymbol{Q}_{\mathrm{cycl},4}$ & $\left(\epsilon,-\epsilon,5-\delta_{\mathrm{cycl}}\right)$ & $\left(0,0,4\right)+\boldsymbol{k}_{\mathrm{cycl},4}$ \\ 
			
			$\boldsymbol{Q}^{*}_{\mathrm{cycl},1}$ &$\left(-\epsilon,-\epsilon,5+\delta_{\mathrm{cycl}}\right)$ & $\left(0,0,6\right)-\boldsymbol{k}_{\mathrm{cycl},1}$\\
			$\boldsymbol{Q}^{*}_{\mathrm{cycl},2}$ &$\left(\epsilon,\epsilon,5+\delta_{\mathrm{cycl}}\right)$& $\left(0,0,6\right)-\boldsymbol{k}_{\mathrm{cycl},2}$\\
			$\boldsymbol{Q}^{*}_{\mathrm{cycl},3}$&$\left(\epsilon,-\epsilon,5+\delta_{\mathrm{cycl}}\right)$& $\left(0,0,6\right)-\boldsymbol{k}_{\mathrm{cycl},3}$\\
			$\boldsymbol{Q}^{*}_{\mathrm{cycl},4}$&$\left(-\epsilon,\epsilon,5+\delta_{\mathrm{cycl}}\right)$& $\left(0,0,6\right)-\boldsymbol{k}_{\mathrm{cycl},4}$\\
			
			$\boldsymbol{Q}_{\mathrm{con},1}$ & $\left(\epsilon,\epsilon,5-\delta_{\mathrm{con}}\right)$  & $\left(0,0,4\right)+\boldsymbol{k}_{\mathrm{con},1}$ \\
			$\boldsymbol{Q}_{\mathrm{con},2}$ & $\left(-\epsilon,-\epsilon,5-\delta_{\mathrm{con}}\right)$ & $\left(0,0,4\right)+\boldsymbol{k}_{\mathrm{con},2}$  \\
			
			$\boldsymbol{Q}^{*}_{\mathrm{con},1}$ &$\left(-\epsilon,-\epsilon,5+\delta_{\mathrm{con}}\right)$   & $\left(0,0,6\right)-\boldsymbol{k}_{\mathrm{con},1}$  \\
			$\boldsymbol{Q}^{*}_{\mathrm{con},2}$ &$\left(\epsilon,\epsilon,5+\delta_{\mathrm{con}}\right)$  & $\left(0,0,6\right)-\boldsymbol{k}_{\mathrm{con},2}$ \\
			$\boldsymbol{Q}_{\mathrm{com}}$&$\left(0,0,5\right)$  &  $\left(0,0,4\right)+\boldsymbol{k}_{\mathrm{com}}$     \\
			$\boldsymbol{Q}_{\mathrm{fan}}$&$\left(0,0,5-\delta_{\mathrm{fan}}\right)$ & $\left(0,0,4\right)+\boldsymbol{k}_{\mathrm{fan}}$ \\
			$\boldsymbol{Q}^{\mathrm{*}}_{\mathrm{fan}}$&$\left(0,0,5+\delta_{\mathrm{fan}}\right)$ & $\left(0,0,6\right)-\boldsymbol{k}_{\mathrm{fan}}$ 
		\end{tabular}
	\end{ruledtabular}
\end{table}

\clearpage

\section{Dependence of magnetic order on the temperature-and-field history}

Between increasing and decreasing magnetic fields, pronounced hysteresis is observed. In the cycloidal and conical phases, distinct changes of the domain populations indicate the single-$k$ nature of the magnetic structures in these phases, as elaborated on in the following.

\begin{figure}
	\includegraphics{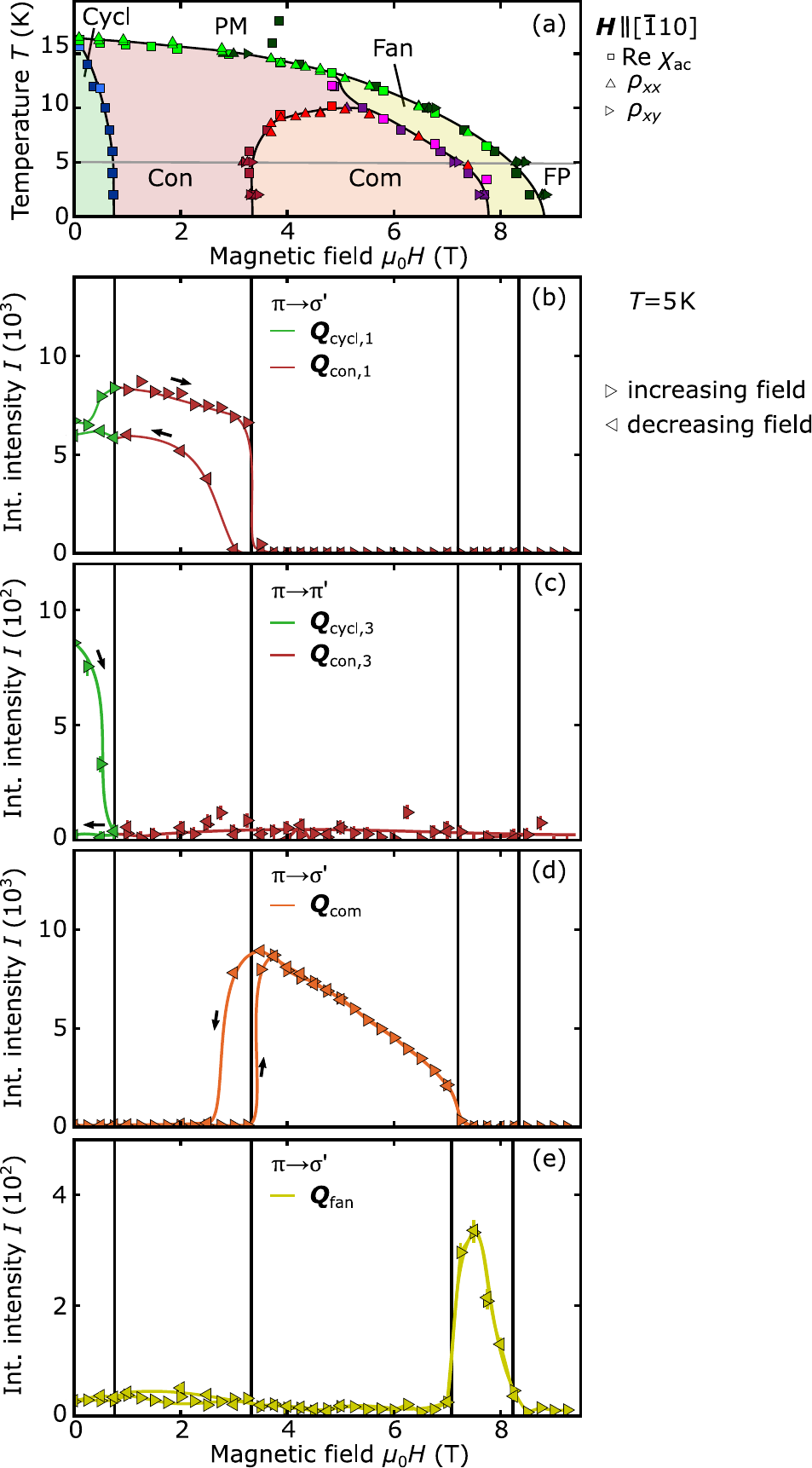}
	\caption{\label{figureS4}FLPA scattering channel-dependent REXS intensity for different field histories. (a)~Magnetic phase diagram of EuPtSi$_{3}$ as shown in Fig.~1(c) of the main text. After initial zero-field cooling to 5~K (cf.\ gray line), REXS was carried out under increasing field from $0~\mathrm{T} \rightarrow 9.25~\mathrm{T}$ and subsequently under decreasing field from $9.25~\mathrm{T} \rightarrow 0~\mathrm{T}$. \mbox{(b)--(e)}~Integrated intensities of magnetic Bragg peaks as a function of field, as inferred from $l$ scans through the corresponding magnetic Bragg peaks by means of numerical integration. Black arrows mark the sweep direction of the magnetic field. The vertical lines indicate phase transitions, as observed after zero-field cooling and subsequent increase of magnetic field.}
\end{figure}

As a point of reference, the magnetic phase diagram of EuPtSi$_{3}$ for field parallel to $\left[\bar{1}10\right]$ is depicted in Fig.~\ref{figureS4}(a). The integrated intensities of the magnetic Bragg peaks characteristic of the cycloidal, conical, commensurate, and fan-like phases are shown in Figs.~\ref{figureS4}(b) to \ref{figureS4}(e) as measured both under increasing and decreasing fields (cf.\ black arrows). In Figs.~\ref{figureS4}(b) and \ref{figureS4}(c), the intensity of two wave vectors, namely one wave vector with an in-plane component perpendicular to the field direction, $\boldsymbol{Q}_{1}$, and one wave vector with an in-plane component parallel to the field direction, $\boldsymbol{Q}_{3}$. The color indicates the respective magnetic phase.

Under increasing field, the intensity at $\boldsymbol{Q}_{1}$ slightly increases, while the intensity at $\boldsymbol{Q}_{3}$ distinctly decreases and vanishes as the conical state emerges. In the commensurate and fan-like phases no intensity is observed at $\boldsymbol{Q}_{1}$  or $\boldsymbol{Q}_{3}$. Under decreasing field, the intensity at $\boldsymbol{Q}_{1}$ is recovered in both the conical and the cycloidal phase. In contrast, no scattering intensity is observed at $\boldsymbol{Q}_{3}$ even in zero field. This finding indicates that the domain population in the cycloidal phase depends on the temperature-and-field history, where magnetic field favors domains with wave vectors having an in-plane component perpendicular to the field direction. This observation is also in excellent agreement with ac susceptibility data~\cite{2022_Bauer_PhysRevMaterials}. Moreover, the dependence of the intensity distributions on the field history is characteristic of single-$k$ magnetic order rather than multi-$k$ order.

In the commensurate and fan-like phases, single-$k$ magnetic structures are observed for which the crystal structure implies only one possible domain. Consistent with this conjecture, as shown in Figs.~\ref{figureS4}(d) and \ref{figureS4}(e), the integrated intensities of magnetic Bragg peaks at $\boldsymbol{Q}_{\mathrm{com}}$ and $\boldsymbol{Q}_{\mathrm{fan}}$ are almost independent of the field history. However, the transition field between the conical and commensurate phase, also referred to as $H_{2}$, clearly differs between increasing and decreasing field values. This discrepancy is observed in both Figs.~\ref{figureS4}(b) and \ref{figureS4}(d) and consistent with the results of ac susceptibility measurements~\cite{2022_Bauer_PhysRevMaterials}. 


\clearpage

\section{Polarization analysis and magnetic structure determination}

As one of the key results of our study, in the following the determination of magnetic structures in the four long-range ordered phases is elaborated on using the full linear polarization analysis (FLPA) technique on REXS data for field along $\left[\bar{1}10\right]$. After a FLPA of the direct X-ray beam, the presentation continues with exemplary data of the FLPAs for the cycloidal, the conical, the commensurate, and the fan-like phase, where a detailed account is provided on the fitting of the experimental data with specific magnetic structures. The section concludes with a discussion of the real-space depictions of the resulting magnetic structures.

The experimental setup used for a FLPA is illustrated in Fig.~\ref{figureS1}. The angles characterizing the direction of the incident and of the scattered beam polarization are denoted $\eta$ and $\eta'$. In order to avoid confusion, the angle along that the scattered polarization is analyzed is denoted $\nu'$, contrasting many studies in literature which refer to this angle as $\eta'$. For further details on the FLPA technique, we refer for instance to Refs.~\cite{2007_Mazzoli_PhysRevB, 2008_Johnson_PhysRevB, 2009_Hatton_JournalofMagnetismandMagneticMaterials, 2012_Shukla_PhysRevB, 2018_Rahn_PhysRevB}.

\subsection{Polarization analysis of the direct beam}
\label{section:FLPAdirectbeam}
At first, a FLPA of the direct beam is carried out in order to determine the degree of linear X-ray polarization reached in the experiment. Subsequently, the resulting Poincar\'{e}--Stokes parameters of the direct beam are also used for optimizing the fitting routine of the magnetic structures. 

\begin{figure}
	\includegraphics{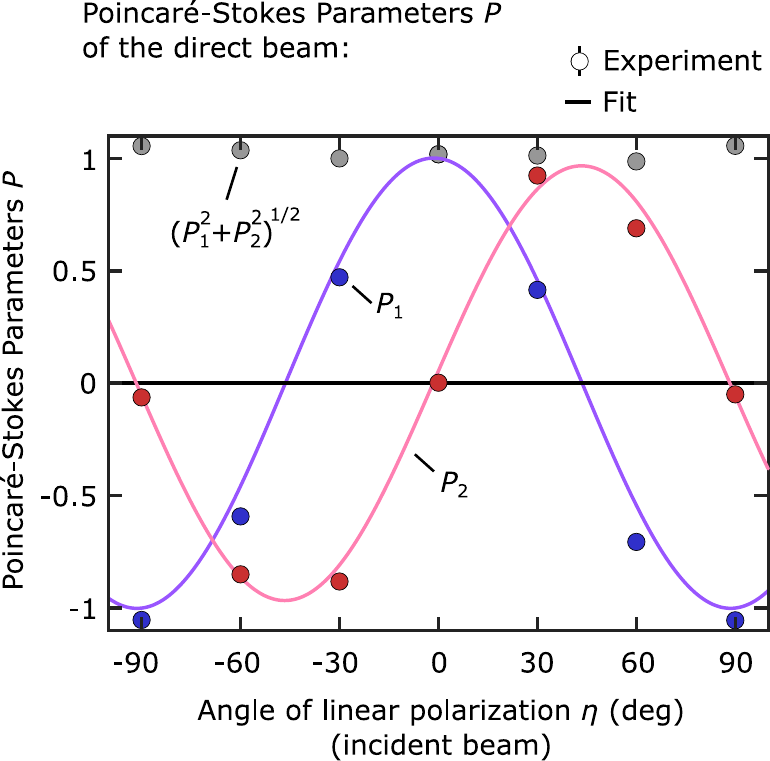}
	\caption{\label{figureS5}Full linear polarization analysis of the incident X-ray beam. Shown are the Poincar\'{e}--Stokes parameters $P_1$ (red symbols) and $P_2$ (blue symbols) as a function of incident polarization angle $\eta$. Circles represent data points as inferred from the experiment. The degree of linear polarization is indicated by the parameter $\sqrt{P_1^2+P_2^2}$ (gray symbols). The solid curves are fits to the experimental data following the relation $C_1\cdot \cos(2\eta)+C_2\cdot \sin(2\eta)$.}
\end{figure}

The Poincar\'{e}--Stokes parameters $P_1$ and $P_2$ of the direct beam are inferred, as described in the main text. As shown in Fig.~\ref{figureS5}, the degree of incident linear X-ray polarization is close to $100\%$ for each incoming polarization angle $\eta$.

To provide an interpolation of $P_1$ and $P_2$ to angles $\eta$, where no experimental data are available, the experimental values of $P_1$ and $P_2$ as a function of $\eta$ are fitted by the superposition of $\cos(2\eta)$ and $\sin(2\eta)$: 
\begin{align}
	\boldsymbol{P}=\left(P_1,P_2,P_3\right)=\left(\cos\left(2\eta\right)-0.049\cdot\sin\left(2\eta\right),\sin\left(2\eta\right)-0.054\cdot\cos\left(2\eta\right),0\right) .   \label{equation:DirectBeamPolarizationFit} 
\end{align}
The fit is used for the calculations presented in Figs.~\ref{figureS7}, \ref{figureS8}, \ref{figureS9}, and \ref{figureS10}.

Even compared to a perfectly polarized X-ray beam and an ideal analyzer, as presented in Eq.~\eqref{equation:PoincareStokesIncidentIdeal}, the deviations are small only, indicating a very high degree of incident linear polarization. Integrated intensities recorded by means of analyzer rocking scans in the $\pi\rightarrow\pi'$ channel and in the $\pi\rightarrow\sigma'$ channel display a ratio of 0.0016, consistent with negligibly small analyzer leakage.

\subsection{FLPAs in the cycloidal, conical, commensurate, and fan-like phases}

As already addressed in the main text, the magnetic structures in the cycloidal, conical, commensurate, and fan-like phases for field along $\left[\bar{1}10\right]$ were determined by means of FLPAs on the magnetic Bragg peaks at $\boldsymbol{Q}_{\mathrm{cycl},1}$, $\boldsymbol{Q}_{\mathrm{con},1}$, $\boldsymbol{Q}_{\mathrm{com}}$, and $\boldsymbol{Q}_{\mathrm{fan}}$, respectively. For each FLPA, 7 different angles $\eta_i$ ($1\leq i\leq 7$) were chosen for the incident linear polarization (the same angles were chosen also for the FLPA of the direct beam). In turn, for each incident polarization, the scattered beam was analyzed along 9 different directions in order to determine the Poincar\'{e}--Stokes parameters $P'_1$ and $P'_2$ of the scattered beam. Integrated intensities of the Bragg peaks in each polarization channel were inferred from a rocking scan of the analyzer crystal~\cite{2004_Detlefs_PhysicaBCondensedMatter}.

\begin{figure}
	\includegraphics{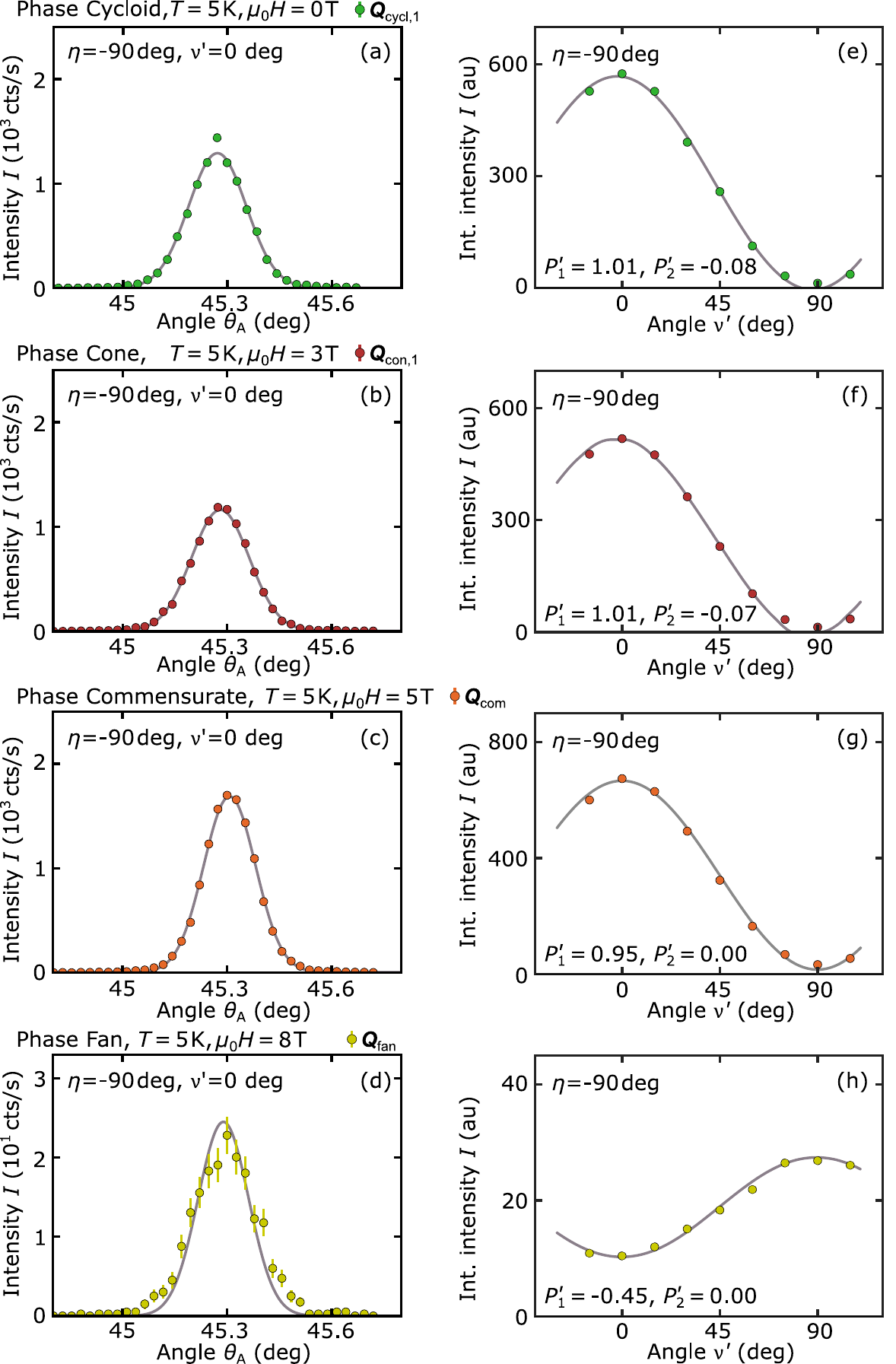}
	\caption{\label{figureS6}Typical data from the FLPAs in the cycloidal, conical, commensurate, and fan-like phases for magnetic field applied along $\left[\bar{1}10\right]$. \mbox{(a)--(d)}~Rocking scans of the analyzer crystal in the polarization channel $\pi\rightarrow\sigma'$. The curves were fitted with Gaussian profiles and integrated. \mbox{(e)--(h)}~Integrated intensities as a function of the analyzer angle $\nu'$ at an incident polarization $\pi$. Fitting the integrated intensities as a function of $\nu'$ yields the Poincar\'{e}--Stokes parameters of the scattered beam as described in the text.}
\end{figure}

Typical data belonging to the FLPAs in the four long-range ordered phases are presented in Fig.~\ref{figureS6}. Rocking scans of the analyzer crystals around the Bragg peaks ($\boldsymbol{Q}_{\mathrm{cycl},1}$, $\boldsymbol{Q}_{\mathrm{con},1}$, $\boldsymbol{Q}_{\mathrm{com}}$, and $\boldsymbol{Q}_{\mathrm{fan}}$) in the polarization channel $\pi\rightarrow\sigma'$ are shown in Figs.~\ref{figureS6}(a) to \ref{figureS6}(d). The integrated intensities of these four Bragg peaks as a function of the analyzer angle $\nu'$ at an incident polarization $\pi$ is subsequently shown in Figs.~\ref{figureS6}(e) to \ref{figureS6}(h). Using the relation
\begin{align}
	I\left(\nu'\right)=I_0\cdot S+I_0\cdot \left[ P'_1\cdot \cos\left(2\nu'\right)+P'_2\cdot \sin\left(2\nu'\right)\right] \, ,
\end{align}
where $S=(1+\cos^2(2\theta_{\mathrm{A}}))/\sin^2(2\theta_{\mathrm{A}})$ represents the spillover parameter, the Poincar\'{e}--Stokes parameters of the scattered beam were inferred by means of fitting the integrated intensities as a function of $\nu'$. For our experiments, $S$ is essentially one, $|S-1|\lessapprox 10^{-4}$.

\subsection{Magnetic structure determination}

For each of the four long-range ordered phases, the magnetic structure providing the best fit to the experimental polarization data was determined as follows. First, for each phase, the irreducible representations (IRs) are determined that allow finite magnetic moments, as summarized in Tab.~\ref{tableS3}. For each potential magnetic structure $\boldsymbol{m}$, Poincar\'{e}--Stokes parameters $P'^{\mathrm{calc}}_1\left(\eta_i\right)$ and $P'^{\mathrm{calc}}_2\left(\eta_i\right)$ were calculated using Eq.~\eqref{equation:poincarestokesofscatteredbeam} and assuming that the incident polarization $\boldsymbol{P}$ corresponds to the experimentally determined Poincar\'{e}--Stokes parameters of the direct beam as shown in Fig.~\ref{figureS5}.

Next, the 7 experimentally determined pairs of Poincar\'{e}--Stokes parameters $P'^{\mathrm{exp}}_1\left(\eta_i\right)$ and $P'^{\mathrm{exp}}_2\left(\eta_i\right)$ were compared with the calculated values for each potential magnetic structure by means of least-squares statistics using
\begin{align}
	\chi^2=\sum_{1\leq i\leq7} \frac{\left(   P'^{\mathrm{calc}}_1\left(\eta_i\right) -P'^{\mathrm{exp}}_1\left(\eta_i\right) \right)^2}{\left(\sigma_{P'_1}\left(\eta_i\right)\right)^2}+\sum_{1\leq i\leq7} \frac{\left(   P'^{\mathrm{calc}}_2\left(\eta_i\right) -P'^{\mathrm{exp}}_2\left(\eta_i\right) \right)^2}{\left(\sigma_{P'_2}\left(\eta_i\right)\right)^2}    \, .
\end{align}
Here, $\sigma_{P'_1}\left(\eta_i\right)$ and $\sigma_{P'_2}\left(\eta_i\right)$ denote the statistical errors of the Poincar\'{e}--Stokes parameters inferred from the experiments. The smallest value of $\chi_{\nu}^2=\frac{\chi^2}{N_{\mathrm{o}}-N_{\mathrm{f}}}$, with $N_{\mathrm{o}}$ and $N_{\mathrm{f}}$ denoting the number of observed and fitted parameters, indicated the magnetic structure providing the best fit to the experimental data.

\begin{table}
	\caption{\label{tableS3}Overview of irreducible representations (IRs) of the magnetic Eu site in the space group $I4mm$ for the magnetic propagation vectors $\boldsymbol{k}_{\mathrm{cycl}}$, $\boldsymbol{k}_{\mathrm{con}}$, $\boldsymbol{k}_{\mathrm{com}}$, and $\boldsymbol{k}_{\mathrm{fan}}$. IRs permitting finite magnetic moments are shown with the respective basis functions. The components of the basis vectors (BV) refer to the conventional tetragonal basis. The analysis was carried out by means of the software SARA\textit{h}~\cite{2000_Wills_PhysicaBCondensedMatter}.}
	\begin{ruledtabular}
		\begin{tabular}{c|cc|ccc}
			$\boldsymbol{k}$-vector&  IR  &  BV  &   \multicolumn{3}{c}{BV components}\\
			& &                   &$m_{1}$ & $m_{2}$ & $m_{3}$\\
			\hline
			&$\Gamma_{1}$ & $\Psi_{1}$&        1 &      -1 &     0       \\
			$\boldsymbol{k}_{\mathrm{cycl}}$,$\boldsymbol{k}_{\mathrm{con}}$ &$\Gamma_{2}$ &  $\Psi_{2}$&       1 &      1 &      0    \\
			&   $\Gamma_{2}$        &   $\Psi_{3}$&   0 &     0 &      2     \\
			\hline
			&$\Gamma_{2}$ & $\Psi_{1}$&        0 &      0 &     8   \\
			$\boldsymbol{k}_{\mathrm{com}}$&$\Gamma_{5}$ &  $\Psi_{2}$&       4 &      0 &      0   \\
			&             &   $\Psi_{3}$&   0 &     -4 &      0   \\
			\hline
			&$\Gamma_{2}$ & $\Psi_{1}$&        0 &      0 &     8 \\
			$\boldsymbol{k}_{\mathrm{fan}}$&$\Gamma_{5}$ &  $\Psi_{2}$&       4 &      0 &      0   \\
			&     $\Gamma_{5}$        &   $\Psi_{3}$&   0 &     -4 &      0  \\
		\end{tabular}
	\end{ruledtabular}
\end{table}

For each phase, first the scattering amplitude in Eq.~\eqref{equation:resonantdipolescatteringamplitude} is considered for purely magnetic scattering by means of fixing $D_1=\rho(\bm{k}) F^{\left(0\right)}=0$, thereby assuming that contributions from charge scattering are negligible and not considering contributions from the mixed term setting $D_2\sim F^{\left(1\right)}\cdot K=1$ and  $D_3\sim F^{\left(2\right)}\cdot C\cdot M_0=0$. In the cycloidal, conical, and commensurate phases, one IR fits the data distinctly better than the others. In a second step, the refinement is repeated for this specific IR under consideration of contributions from charge scattering due to a charge density that has the same wave-vector $\bm{k}$ as the magnetic structure by means of allowing for finite values for $D_1=\rho(\bm{k}) F^{\left(0\right)}$ and setting $D_2=F^{\left(1\right)}\cdot K=1$ (where $K$ is a constant that comprises further factors like the modulus of the magnetization).  In the fan-like phase, no IR is in good agreement with the experimental data in the first step and all IRs are reconsidered in the second step, where contributions from charge scattering are included. 

Notably, in phases Commensurate and Fan the fitting procedure is significantly improved, when charge-corrections to the scattering amplitude are included. The presence of charge scattering at magnetic Bragg positions may indicate structural symmetry lowering due to charge-modulations with a modulation period that is equal to that of the magnetic modulation or it may arise from diffuse charge-scattering. 

In a third step, we consider also the third term in the scattering amplitude given in Eq.~\eqref{equation:resonantdipolescatteringamplitude} by considering finite values of $D_3=F^{\left(2\right)}\cdot C\cdot M_0$, where $M_0$ denotes the field-dependent magnetization along $\bm{u}_2$. The respective mixed term of the scattering amplitude that leads to magnetic intensity at the magnetic Bragg peak position at finite magnetic field is given by:
\begin{align}
D_3\cdot \begin{pmatrix}
		2z_2  &  - z_1\cdot\sin\left(\theta\right) + z_3\cdot\cos\left(\theta\right) \\
		z_1\cdot\sin\left(\theta\right)+z_3\cdot\cos\left(\theta\right)   & 0\end{pmatrix}    \, .
\end{align}

Including this mixed term did not considerably improve the fit results. Similarly, fits incorporating this mixed third term but not charge contributions ($D_1=\rho F^{\left(0\right)}=0$) in the scattering amplitude did not provide better results than the fits without charge scattering and without the mixed term.

Experimentally, antiferromagnetic propagation vectors of the form $\left(0.007,0.007,0.927\right)$, $\left(0.007,0.007,0.927\right)$, $\left(0,0,1\right)$, and $\left(0,0,0.887\right)$ were observed in the cycloidal, conical, commensurate, and fan-like phases, as denoted in the conventional tetragonal basis. Typically, group-theoretical considerations and translational properties of magnetic structures are discussed in the primitive basis, in which these four wave vectors read $\left(0.464,0.464,-0.457\right)$, $\left(0.464,0.464,-0.457\right)$, $\left(\frac{1}{2},\frac{1}{2},-\frac{1}{2}\right)$, and $\left(0.444,0.444,-0.444\right)$. In the following, however, similar to the diffraction data in the main text, the magnetic structures are presented in the conventional frame.

In the cycloidal, conical, and fan-like phases, where the magnetic structures possess an incommensurate antiferromagnetic propagation vector $\bm{k}_{\mathrm{ic}}$, the Fourier transform of the magnetic ground state (GS) is given by:

\begin{align}
\bm{M}_{\mathrm{GS}}(\bm{Q})= M_0\cdot \bm{u}_2  \cdot \delta(\bm{Q}+\bm{G})+ \bm{M}(\bm{k}_{\mathrm{ic}})\cdot  \delta(\bm{Q}-\bm{k}_{\mathrm{ic}}+\bm{G}) + \bm{M}(-\bm{k}_{\mathrm{ic}})\cdot \delta(\bm{Q}+\bm{k}_{\mathrm{ic}}+\bm{G}) \,
\end{align}
where $M_0$ denotes the field-induced net magnetization. Further, we considered the charge density $\rho(\bm{Q})=\rho_0 \cdot \delta(\bm{Q}+\bm{G}+\bm{k}_{\mathrm{ic}})+\rho_0 \cdot \delta(\bm{Q}+\bm{G}-\bm{k}_{\mathrm{ic}})$.

In the commensurate phase, the magnetic structure possesses a commensurate antiferromagnetic wave-vector $\bm{k}_{\mathrm{com}}$ and the Fourier transform is given by:
\begin{align}
\bm{M}_{\mathrm{GS}}(\bm{Q})= M_0\cdot \bm{u}_2 \cdot \delta(\bm{Q}+\bm{G})+   \bm{M}(\bm{k}_{\mathrm{com}}) \cdot \delta(\bm{Q}-\bm{k}_{\mathrm{com}}+\bm{G}) 
\end{align}
In addition, we considered the charge density $\rho(\bm{Q})=\rho_0 \cdot \delta(\bm{Q}+\bm{G}-\bm{k}_{\mathrm{com}})$.

\subsubsection{Cycloidal phase}

The presentation of our structure determination starts with the cycloidal phase assuming negligible contribution due to charge scattering, $D_1=0$. In order to probe the IR $\Gamma_1$, Fourier-transformed magnetic structures were considered with $\boldsymbol{M}\left(\bm{k}_{\mathrm{cycl}}\right)=\left(1+i\cdot C_1\right)\frac{\Psi_1}{\sqrt{2}}$ and $C_1$ being a real number. 
At the minimum, the goodness of the fit corresponds to $\chi^2=63164$ and $\chi^2_{\nu}=\frac{\chi^2}{N_{\mathrm{o}}-N_{\mathrm{f}}}=\frac{\chi^2}{14-1}\approx4859$ ($N_{\mathrm{o}}=14$ and $N_{\mathrm{f}}=1$ denote the number of observed and fitted parameters). The relatively large values indicate an insufficient agreement between experiment and the $\Gamma_1$ representation.

In order to probe the IR $\Gamma_2$, Fourier-transformed magnetic structures were considered with $\boldsymbol{M}\left(\bm{k}_{\mathrm{cycl}}\right)= \frac{1}{2}\cdot \frac{\Psi_2}{\sqrt{2}}+\frac{1}{2\cdot\mathrm{i}}\cdot C_1\cdot \frac{\Psi_3}{2}+C_2\cdot \frac{\Psi_3}{2}$. 
Two minima of $\chi^2$ were identified, where the goodness of the fit corresponds to $\chi^2=282.7$ and $\chi^2_{\nu}=\frac{\chi^2}{14-2}=23.6$, respectively, indicating good agreement between experiment and calculations. The magnetic structure associated with one of the two minima is given by
\begin{align}
	\boldsymbol{M}_{\mathrm{cycl},1}\left(\bm{k}_{\mathrm{cycl}}\right)=\frac{1}{2}\cdot\left(\frac{1}{\sqrt{2}},\frac{1}{\sqrt{2}},\frac{1}{\mathrm{i}}\cdot 1.56- 0.236 \right) \, .
\end{align}
The components of $\boldsymbol{M}$ refer to the orthonormal basis vectors $\hat{\boldsymbol{a}}_1=\boldsymbol{a}_1/\left|\boldsymbol{a}_1\right|$, $\hat{\boldsymbol{a}}_2=\boldsymbol{a}_2/\left|\boldsymbol{a}_2\right|$, and $\hat{\boldsymbol{a}}_3=\boldsymbol{a}_3/\left|\boldsymbol{a}_3\right|$ directed along the crystallographic directions $\left[100\right]$, $\left[010\right]$, and $\left[001\right]$, respectively. In turn, at the second minimum the magnetic structure is given by
\begin{align}
	\boldsymbol{M}\left(\bm{k}_{\mathrm{cycl}}\right)=\frac{1}{2}\cdot\left(\frac{1}{\sqrt{2}},\frac{1}{\sqrt{2}},-\frac{1}{\mathrm{i}}\cdot 1.56- 0.236 \right) \, .
\end{align}
The two minima correspond to two inverted wave-vectors and therefore to the two different realizations of the noncentrosymmetric crystal structure of EuPtSi$_{3}$.

The real-space description of the corresponding magnetic structure is given by
\begin{align}
	\boldsymbol{m}_{\mathrm{cycl},1}\left(\boldsymbol{R}\right)=& m_0\cdot\frac{1}{\sqrt{2}}\begin{pmatrix}    1 \\ 1 \\ 0 \end{pmatrix} \cdot\cos\left(\boldsymbol{k}_{\mathrm{cycl}}\cdot\boldsymbol{R}\right)+1.56 \cdot m_0\cdot \begin{pmatrix}    0 \\ 0 \\ 1 \end{pmatrix}\cdot  \sin\left(\boldsymbol{k}_{\mathrm{cycl}}\cdot\boldsymbol{R}\right)-  \nonumber\\ &- 0.236\cdot m_0\cdot \begin{pmatrix}    0 \\ 0 \\ 1 \end{pmatrix}\cdot\cos\left(\boldsymbol{k}_{\mathrm{cycl}}\cdot\boldsymbol{R}\right)\,  ,
\end{align}
where $\boldsymbol{R}$ denotes the real-space coordinate vector of a europium position, as described in the conventional basis. 

\begin{figure}
	\includegraphics{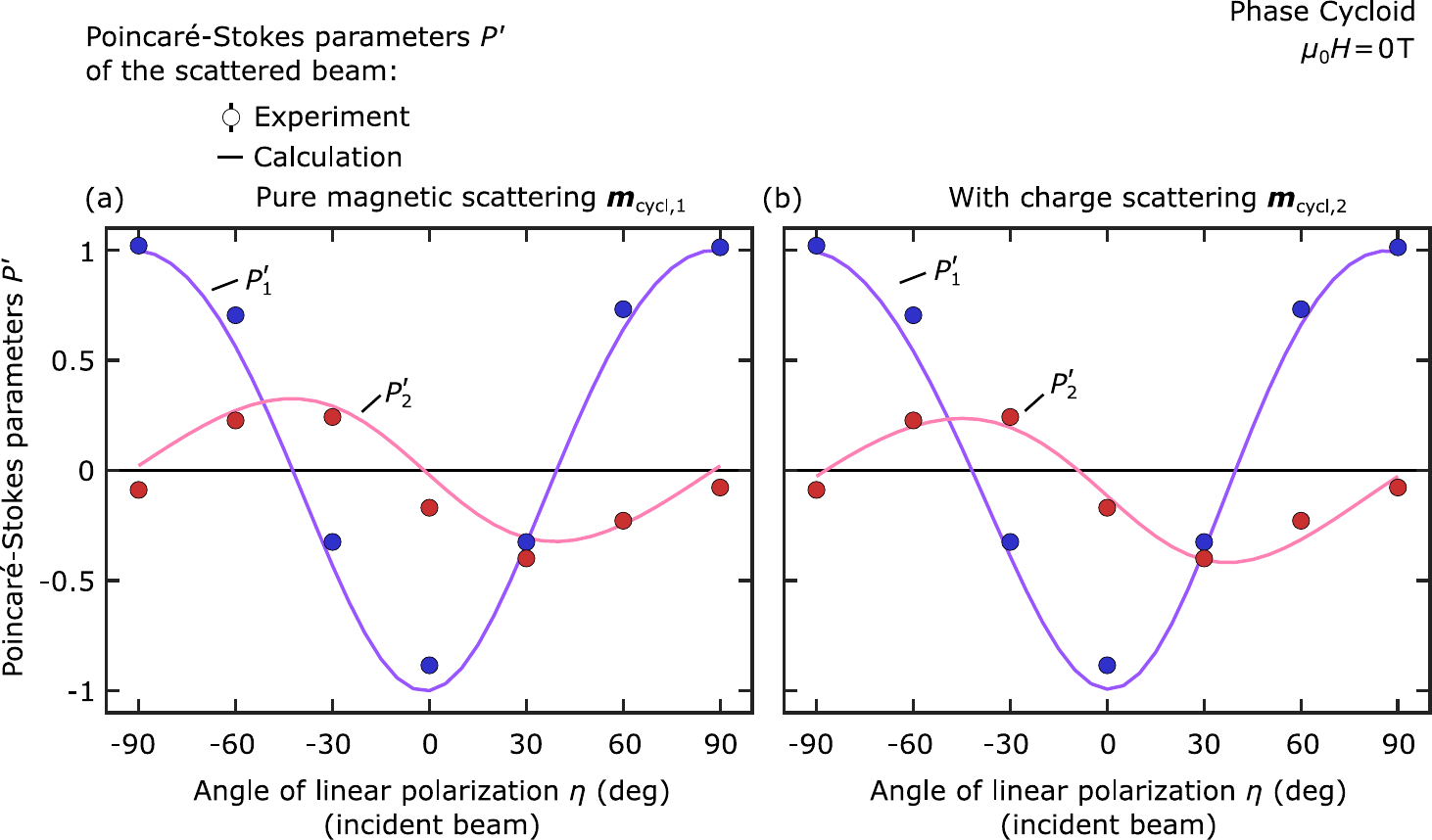}
	\caption{\label{figureS7}Full linear polarization analysis on the magnetic Bragg peak at $\boldsymbol{Q}_{\mathrm{cycl},1}$ in the cycloidal phase. As a function of the angle $\eta$, Poincar\'{e}--Stokes parameters of the scattered beam inferred from the experiments (data points) are compared calculated values (solid lines) for the best-fitting magnetic structures. (a)~Calculations for the structure $\boldsymbol{m}_{\mathrm{cycl},1}$ without contributions from charge scattering. (b)~Calculations for the structure $\boldsymbol{m}_{\mathrm{cycl},2}$ including contributions from charge scattering.}
\end{figure}

In a second step, the fit using the IR $\Gamma_2$ was optimized further by means of also including charge scattering contributions to the scattering amplitude in Eq.~\eqref{equation:resonantdipolescatteringamplitude} by setting $D_2=1$ and allowing for finite values of $D_1$. 
Minimization of $\chi^2$ provided at the minimum $\chi^2=140.47$ and $\chi^2_{\nu}=\chi^2 / \left(14-3\right)=12.77$. The resulting magnetic structure is given by
\begin{align}
	\boldsymbol{M}_{\mathrm{cycl},2}\left(\bm{k}_{\mathrm{cycl}}\right)=\frac{1}{2}\cdot\left(\frac{1}{\sqrt{2}},\frac{1}{\sqrt{2}},\frac{1}{\mathrm{i}}\cdot 1.56- 0.235 \right) \,
\end{align}
at a charge scattering parameter of $D_1=-0.04$. The real-space description of this magnetic structure reads
\begin{align}
	\boldsymbol{m}_{\mathrm{cycl},2}\left(\boldsymbol{R}\right)=& m_0\cdot\frac{1}{\sqrt{2}}\begin{pmatrix}    1 \\ 1 \\ 0 \end{pmatrix} \cdot\cos\left(\boldsymbol{k}_{\mathrm{cycl}}\cdot\boldsymbol{R}\right)+1.56 \cdot m_0\cdot \begin{pmatrix}    0 \\ 0 \\ 1 \end{pmatrix} \cdot\sin\left(\boldsymbol{k}_{\mathrm{cycl}}\cdot\boldsymbol{R}\right)-  \nonumber\\ &- 0.235\cdot m_0\cdot \begin{pmatrix}    0 \\ 0 \\ 1 \end{pmatrix}\cdot\cos\left(\boldsymbol{k}_{\mathrm{cycl}}\cdot\boldsymbol{R}\right)  \,   .
\end{align}

In Figs.~\ref{figureS7}(a) and \ref{figureS7}(b), the Poincar\'{e}--Stokes parameters inferred from the experiments (data points) are compared with those calculated from the structures $\boldsymbol{m}_{\mathrm{cycl},1}$ and $\boldsymbol{m}_{\mathrm{cycl},2}$ (solid lines). For the visualization of these calculations in terms of solid curves, the fits shown in Fig.~\ref{figureS5} were used as the incident polarization.

Although the third term in Eq.~\eqref{equation:resonantdipolescatteringamplitude} does not contribute to the scattering due to the absence of field-induced net magnetization, we repeat the fitting procedure allowing for finite values of $D_3$ as a consistency check. The best fit was obtained for $D_3=0.000$

\subsubsection{Conical phase}

In the conical phase, the magnetic structure corresponds to a superposition of an antiferromagnetic modulation with propagation vector $\boldsymbol{k}_{\mathrm{con}}$ and a ferromagnetic contribution with propagation vector $\boldsymbol{k}=0$, the latter of which results in an uniform magnetization component along the field direction $\left[\bar{1}10\right]$. In our REXS study, only the antiferromagnetic modulation was investigated by a FLPA on the Bragg peak at $\boldsymbol{Q}_{\mathrm{con,1}}$. 

Following the same procedure as for the cycloidal phase, in a first step each of the IRs that permits finite magnetic moments was probed, considering purely magnetic scattering by setting $D_1=0$ in Eq.~\eqref{equation:resonantdipolescatteringamplitude} and by neglecting the last term of the scattering amplitude setting $D_3=0$. For IR $\Gamma_1$ the best fit resulted in the values $\chi^2=37096.7$ and $\chi_{\nu}^2=\chi^2/\left(14-1\right)=2853.6$, while IR $\Gamma_2$ values $\chi^2=238.2$ and $\chi_{\nu}^2=\chi^2/\left(14-2\right)=19.9$ were obtained, clearly identifying the latter IR as the most likely candidate. One of the two structures that minimizes the sum of squared deviations is given by
\begin{align}
	\boldsymbol{M}_{\mathrm{con},1}\left(\bm{k}_{\mathrm{con}}\right)=\frac{1}{2}\cdot\left(\frac{1}{\sqrt{2}},\frac{1}{\sqrt{2}},\frac{1}{\mathrm{i}}\cdot 1.712- 0.222 \right) \, .
\end{align}
In real space, the antiferromagnetically modulated component to the magnetic structure is given by
\begin{align}
	\boldsymbol{m}_{\mathrm{con},1}\left(\boldsymbol{R}\right)=& m_0\cdot\frac{1}{\sqrt{2}}\begin{pmatrix}    1 \\ 1 \\ 0 \end{pmatrix} \cdot\cos\left(\boldsymbol{k}_{\mathrm{con}}\cdot\boldsymbol{R}\right)+1.712 \cdot m_0\cdot \begin{pmatrix}    0 \\ 0 \\ 1 \end{pmatrix} \cdot\sin\left(\boldsymbol{k}_{\mathrm{con}}\cdot\boldsymbol{R}\right)-  \nonumber\\ &- 0.222\cdot m_0\cdot \begin{pmatrix}    0 \\ 0 \\ 1 \end{pmatrix}\cdot\cos\left(\boldsymbol{k}_{\mathrm{con}}\cdot\boldsymbol{R}\right)   \, .
\end{align}
There is also a second minimum corresponding to the inverted crystal structure and given by:
\begin{align}
	\boldsymbol{M}\left(\bm{k}_{\mathrm{con}}\right)=\frac{1}{2}\cdot\left(\frac{1}{\sqrt{2}},\frac{1}{\sqrt{2}},-\frac{1}{\mathrm{i}}\cdot 1.712- 0.222 \right) \, .
\end{align}

\begin{figure}
	\includegraphics{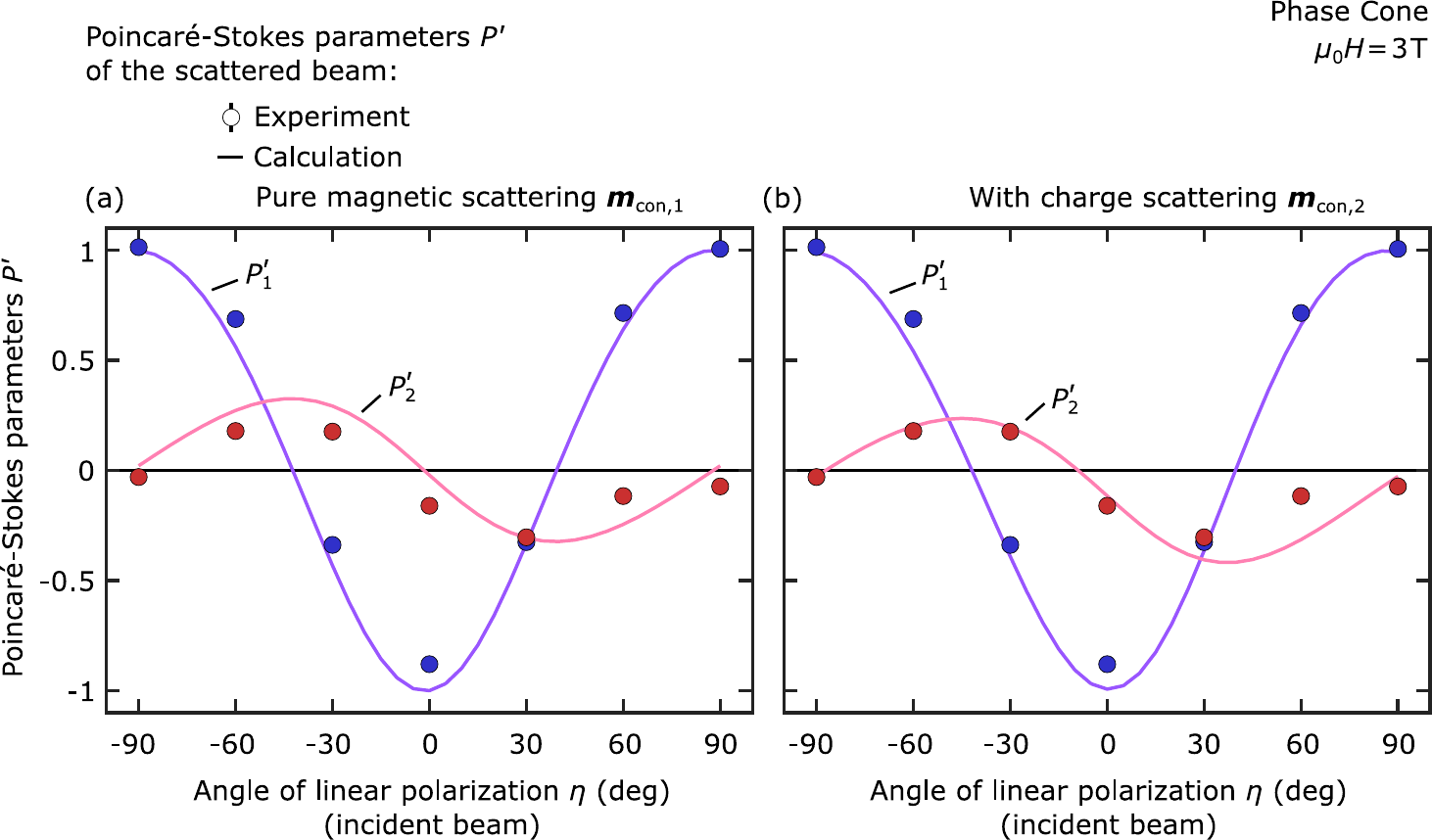}
	\caption{\label{figureS8}Full linear polarization analysis on the magnetic Bragg peak at $\boldsymbol{Q}_{\mathrm{con},1}$ in the conical phase. As a function of the angle $\eta$, Poincar\'{e}--Stokes parameters of the scattered beam inferred from the experiments (data points) are compared calculated values (solid lines) for the best-fitting magnetic structures. (a)~Calculations for the structure $\boldsymbol{m}_{\mathrm{con},1}$ without contributions from charge scattering. (b)~Calculations for the structure $\boldsymbol{m}_{\mathrm{con},2}$ including contributions from charge scattering.}
\end{figure}

In a second step, like for the cycloidal phase, the fit is further improved by including contributions from charge scattering. The best fit resulted in $\chi^2=175.6$ and $\chi_{\nu}^2=\chi^2/\left(14-3\right)=16.0$ and the structure of the first domain that minimizes the sum of squared deviations is given by
\begin{align}
	\boldsymbol{M}_{\mathrm{con},2}\left(\bm{k}_{\mathrm{con}}\right)=\frac{1}{2}\cdot\left(\frac{1}{\sqrt{2}},\frac{1}{\sqrt{2}},\frac{1}{\mathrm{i}}\cdot 1.716- 0.235 \right) \, .
\end{align}
at a charge scattering parameter of $D_1=-0.02$. The real-space description of the antiferromagnetic modulation reads
\begin{align}
	\boldsymbol{m}_{\mathrm{con},2}\left(\boldsymbol{R}\right)=& m_0\cdot\frac{1}{\sqrt{2}}\begin{pmatrix}    1 \\ 1 \\ 0 \end{pmatrix} \cdot\cos\left(\boldsymbol{k}_{\mathrm{con}}\cdot\boldsymbol{R}\right)+1.716 \cdot m_0\cdot \begin{pmatrix}    0 \\ 0 \\ 1 \end{pmatrix}\cdot \sin\left(\boldsymbol{k}_{\mathrm{con}}\cdot\boldsymbol{R}\right)-  \nonumber\\ &- 0.235\cdot m_0\cdot \begin{pmatrix}    0 \\ 0 \\ 1 \end{pmatrix}\cdot\cos\left(\boldsymbol{k}_{\mathrm{con}}\cdot\boldsymbol{R}\right) \,
\end{align}

In Figs.~\ref{figureS8}(a) and \ref{figureS8}(b), the Poincar\'{e}--Stokes parameters from the experiments are compared with those calculated from the structures $\boldsymbol{m}_{\mathrm{con},1}$ and  $\boldsymbol{m}_{\mathrm{con},2}$.

A refinement including also the third term of the scattering amplitude in Eq.~\ref{equation:resonantdipolescatteringamplitude} did not improve the fit. The best fit that was obtained for $D_3=0.000$.

\subsubsection{Commensurate phase}

In the commensurate phase, the magnetic structure is a superposition of an antiferromagnetic modulation with wave-vector $\boldsymbol{k}_{\mathrm{com}}$ and a uniform magnetization component associated with wave vector $\boldsymbol{k}=0$. In our REXS study we studied the antiferromagnetic component by means of a FLPA on the magnetic Bragg peak at $\boldsymbol{Q}_{\mathrm{com}}$.

For the determination of the magnetic structure, again at first purely magnetic scattering is assumed, $D_1=0$, without including the mixed term, $D_3=0$. For the fit with the IR $\Gamma_5$, the Fourier-transformed magnetic structures $\boldsymbol{M}(\bm{k}_{\mathrm{com}})=\left[\Psi_2-\Psi_3+C_1\cdot\left(\Psi_2+\Psi_3\right)\right]\cdot\frac{1}{4}$ were considered. The best fit resulted in $\chi^2=13283$ and $\chi_{\nu}^2=\chi^2/\left(14-1\right)=1021.8$, indicating insufficient agreement between $\Gamma_5$ and experimental data. To probe the IR $\Gamma_2$, the Fourier-transformed magnetic structure $\boldsymbol{M}_{\mathrm{com}}(\bm{k}_{\mathrm{com}})=\frac{\Psi_1}{8}$ was considered. At the minimum, the weighted sum of squared deviations resulted in $\chi^2=374.3$ and $\chi_{\nu}^2=\chi^2/14=26.7$,  indicating relatively good agreement between $\Gamma_2$ and experimental data. 

\begin{figure}
	\includegraphics{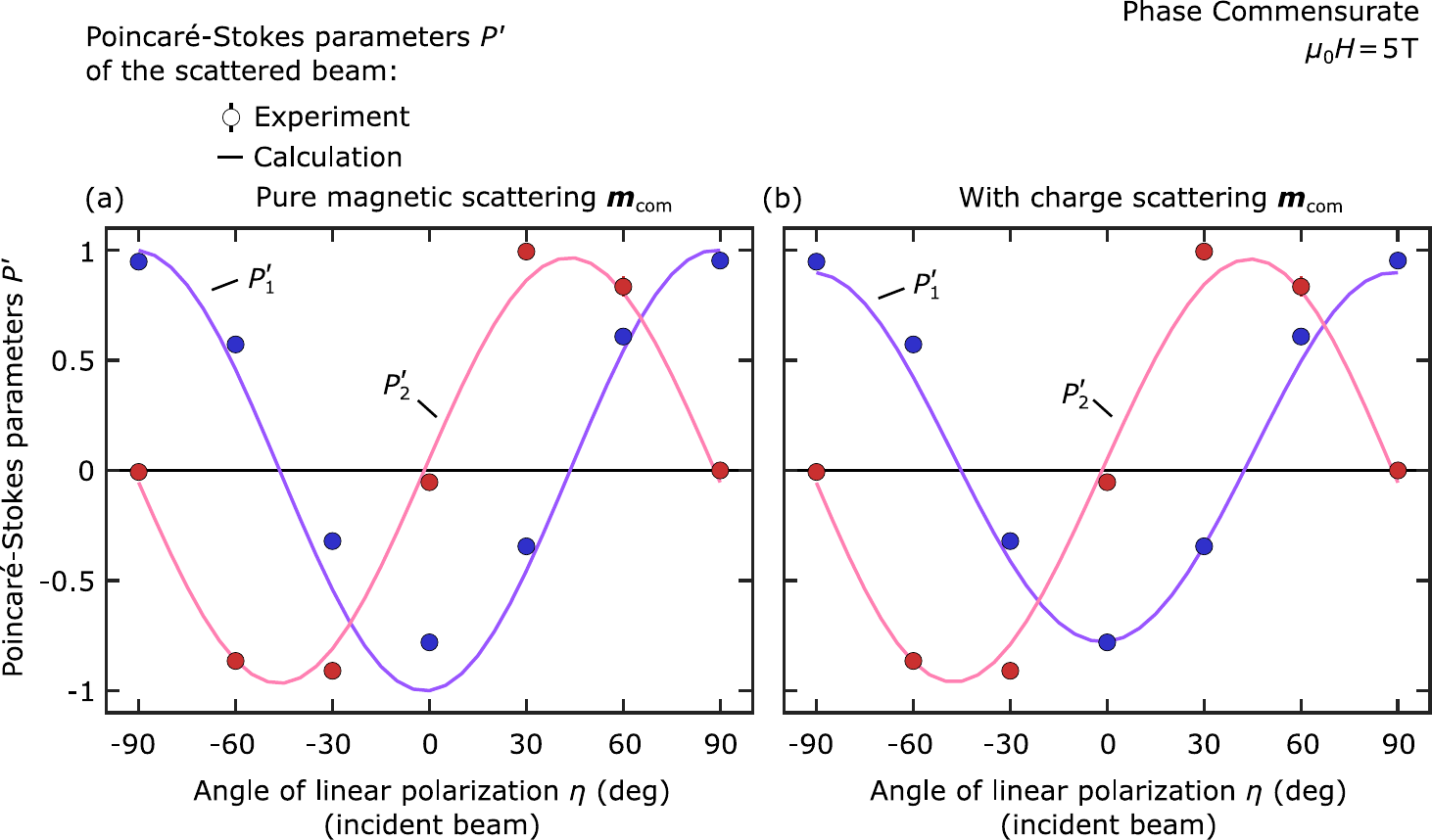}
	\caption{\label{figureS9}Full linear polarization analysis on the magnetic Bragg peak at $\boldsymbol{Q}_{\mathrm{com}}$ in the cycloidal phase. As a function of the angle $\eta$, Poincar\'{e}--Stokes parameters of the scattered beam inferred from the experiments (data points) are compared calculated values (solid lines) for the best-fitting magnetic structures. (a)~Calculations for the structure $\boldsymbol{m}_{\mathrm{com}}$ without contributions from charge scattering. (b)~Calculations for the structure $\boldsymbol{m}_{\mathrm{com}}$ including contributions from charge scattering.}
\end{figure}

In a second step, again contributions from charge scattering were included, resulting in the improved values $\chi^2=49.9$ and $\chi_{\nu}^2=\chi^2/\left(14-1\right)=3.8$ at a charge scattering parameter of $D_1=-0.15$. This value is larger when compared to the situation in the cycloidal and the conical phase. The corresponding magnetic structure of the antiferromagnetic modulation in real-space is given by
\begin{align}
	\boldsymbol{m}_{\mathrm{com}}\left(\boldsymbol{R}\right)=& m_0\cdot\begin{pmatrix}    0 \\ 0 \\ 1 \end{pmatrix} \cdot\cos\left(\boldsymbol{k}_{\mathrm{com}}\cdot\boldsymbol{R}\right)\, .
\end{align}

In Figs.~\ref{figureS9}(a) and \ref{figureS9}(b), the Poincar\'{e}--Stokes parameters from the experiments are compared with those calculated from the structure $\boldsymbol{m}_{\mathrm{com}}$ with and without consideration of contributions from charge scattering. 

A refinement including also the third term of the scattering amplitude in Eq.~\ref{equation:resonantdipolescatteringamplitude} resulted in $D_3=0.015$ and did not significantly improve the fit, as $\chi^2=48.6$ and $\chi_{\nu}^2=\chi^2/\left(14-2\right)=4.1$.

The integrated intensities of the analyzer rocking scans performed for the $\sigma\rightarrow\sigma'$ channel and for the $\pi\rightarrow\sigma'$ channel display a ratio of $0.14$. This value is orders of magnitude larger than the analyzer leakage and hints at strong charge-scattering contributions in the $\sigma\rightarrow\sigma'$ channel.

\subsubsection{Fan-like phase}

In the fan-like phase, the magnetic structure corresponds to a superposition of an antiferromagnetic modulation with propagation vector $\boldsymbol{k}_{\mathrm{fan}}$ and a uniform magnetization component with propagation vector $\boldsymbol{k}=0$. The antiferromagnetic contribution was studied by means of a FLPA on the Bragg peak at $\boldsymbol{Q}_{\mathrm{fan}}$.

Akin to the other three phases, for the FLPA in fan-like phase the experimental data were fitted with all IRs that exhibit finite magnetic moments. In contrast to the previous phases, however, all fits assuming purely magnetic scattering amplitude resulted in relatively bad refinements. To probe the IR $\Gamma_5$, the Fourier transformed magnetic structures $\boldsymbol{M}(\bm{k}_{\mathrm{fan}})=\left[\Psi_2-\Psi_3+C_1\cdot\left(\Psi_2+\Psi_3\right)+\mathrm{i}\cdot C_2\cdot\left(\Psi_2+\Psi_3\right)\right]\cdot\frac{1}{4}$ were considered. 
To probe the IR $\Gamma_2$, the structure $\boldsymbol{M}=\frac{\Psi_1}{8}$ was considered, resulting in values $\chi^2=31811.6$ and $\chi_{\nu}^2=\chi^2/\left(14-2\right)=2272.6$. All these values indicate insufficient agreement between calculations and experiments, which may be attributed to strong contributions from charge scattering.

\begin{figure}
	\includegraphics{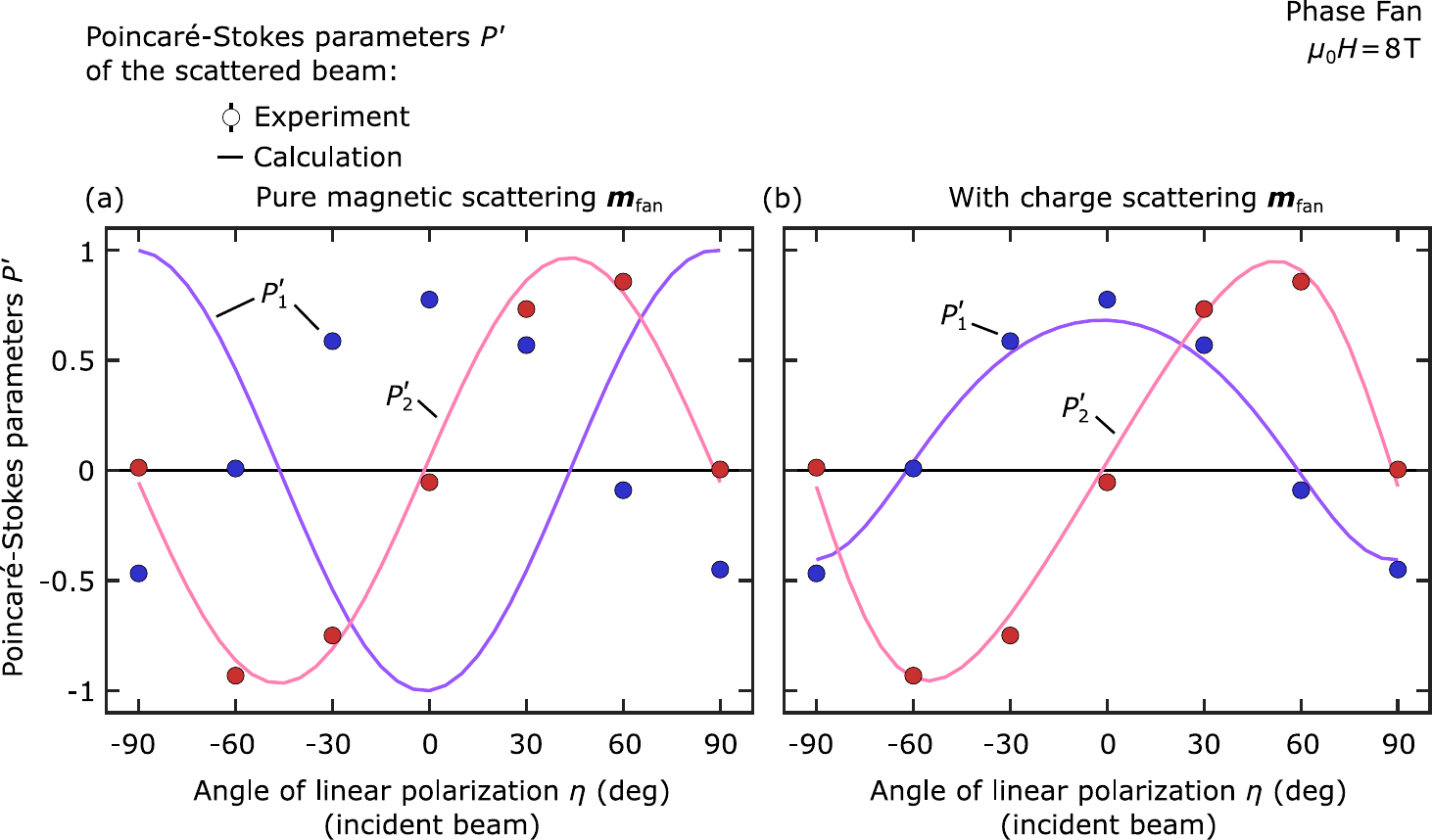}
	\caption{\label{figureS10}Full linear polarization analysis on the magnetic Bragg peak at $\boldsymbol{Q}_{\mathrm{fan}}$ in the cycloidal phase. As a function of the angle $\eta$, Poincar\'{e}--Stokes parameters of the scattered beam inferred from the experiments (data points) are compared calculated values (solid lines) for the best-fitting magnetic structures. (a)~Calculations for the structure $\boldsymbol{m}_{\mathrm{fan}}$ without contributions from charge scattering. (b)~Calculations for the structure $\boldsymbol{m}_{\mathrm{fan}}$ including contributions from charge scattering.}
\end{figure}

Therefore, in contrast to the other three phases, all possible IRs were considered in the second step when allowing for finite contributions from charge scattering. For the IR $\Gamma_5$, the same structure as in the first step was considered. The parameter $D_2$ was set to $1$ and the charge scattering parameter was considered. The best fit resulted in values $\chi^2=2572.9$ and $\chi_{\nu}^2=\chi^2/\left(14-3\right)=233.9$, indicating improved but still insufficient agreement between calculations and experiments. 

For the IR $\Gamma_2$, the same structure as in the first step was considered. At the minimum, the goodness of the fit corresponds to $\chi^2=117.0$ and $\chi_{\nu}^2=\chi^2/\left(14-1\right)=9.0$, indicating excellent agreement between the $\Gamma_2$ representation and the experiments. Here, the size of the parameter $D_1=-0.93$ indicates rather strong contributions from charge scattering. The real-space description of the antiferromagnetic contribution is given by:
\begin{align}
	\boldsymbol{m}_{\mathrm{fan}}\left(\boldsymbol{R}\right)=& m_0\cdot\begin{pmatrix}    0 \\ 0 \\ 1 \end{pmatrix} \cdot\cos\left(\boldsymbol{k}_{\mathrm{fan}}\cdot\boldsymbol{R}\right)\, .
\end{align}

In Figs.~\ref{figureS10}(a) and \ref{figureS10}(b), the Poincar\'{e}--Stokes parameters from the experiments are compared with those calculated from the structure $\boldsymbol{m}_{\mathrm{fan}}$ with and without consideration of contributions from charge scattering.

A fit including also the third term of the scattering amplitude in Eq.~\ref{equation:resonantdipolescatteringamplitude} resulted in $D_3=0.020$. The goodness of the fit essentially remains unchanged, as $\chi^2=106.07$ and $\chi_{\nu}^2=\chi^2/\left(14-2\right)=8.8$. Compared to the commensurate phase, the parameter $D_3$ has increased, which is in agreement with an increase of the net magnetization.

The ratio of integrated intensities of the analyzer rocking scans performed for the $\sigma\rightarrow\sigma'$ channel and for the $\pi\rightarrow\sigma'$ channel exhibit the ratio of $7.02$. This value hints at relatively strong charge-scattering contributions in the $\sigma\rightarrow\sigma'$ channel.

\subsection{Overview of the resulting magnetic structures}

\begin{figure}
	\includegraphics[width=0.9\linewidth]{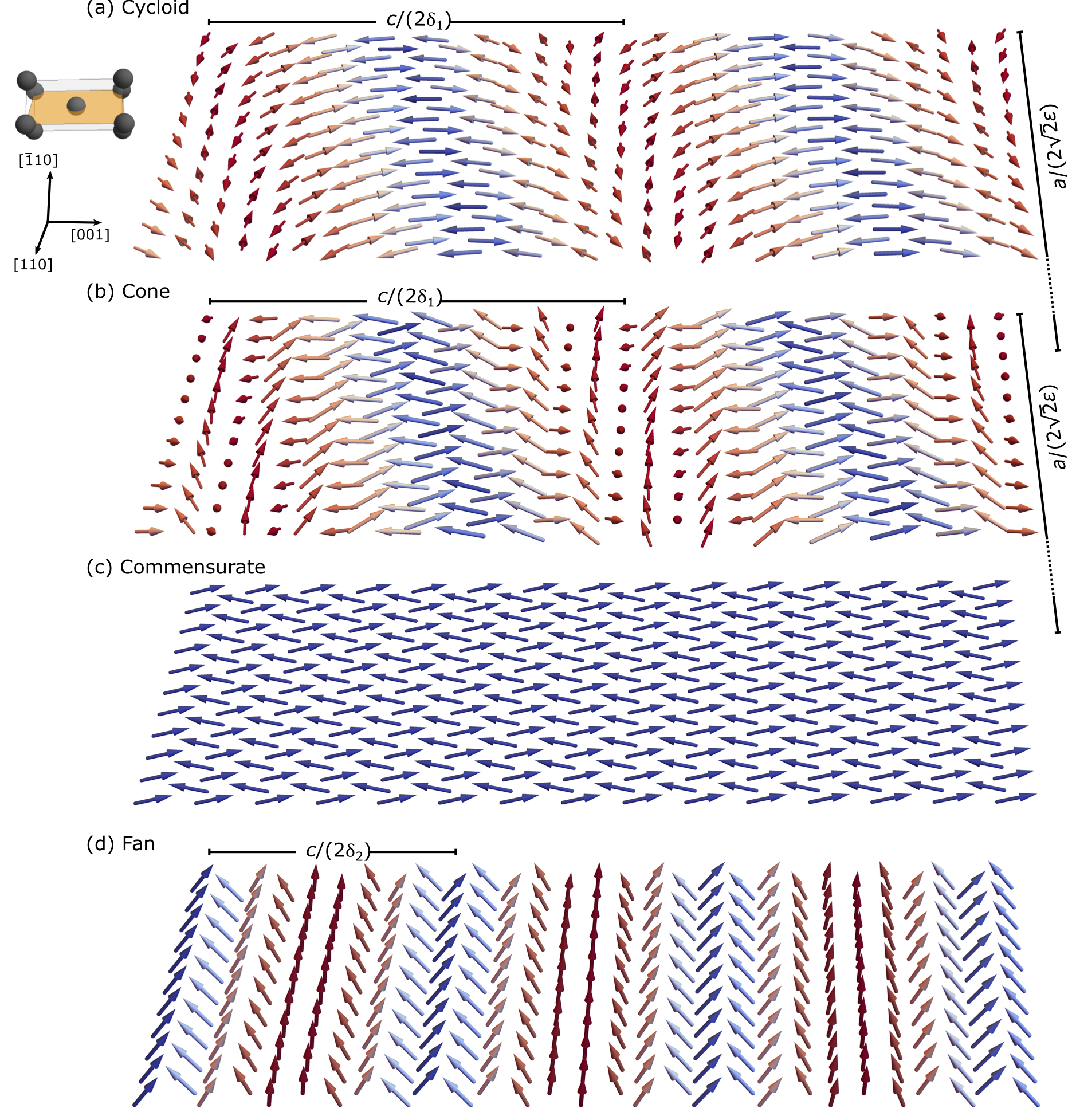}
	\caption{\label{figureS11}Magnetic structures in the cycloidal, conical, commensurate, and fan-like phases as inferred from the FLPAs. The real-space textures are depicted in a $\left(1\bar{1}0\right)$ plane. For the four different phases, one domain of magnetic structure is shown as associated with the Bragg peaks at $\boldsymbol{Q}_{\mathrm{cycl},1}$, $\boldsymbol{Q}_{\mathrm{con},1}$, $\boldsymbol{Q}_{\mathrm{com}}$, and $\boldsymbol{Q}_{\mathrm{fan}}$, respectively.}
\end{figure}

An overview of the magnetic structures resulting from the FLPAs in the four phases is presented in Fig.~\ref{figureS11} and in Fig.~3 of the main text in terms of real-space visualizations of the magnetic textures. Modulation lengths inferred from the data in both field configurations, as presented in the main text and in the following, agree with each other within the error margins.

In the cycloidal phase at zero magnetic field, the magnetic structure is an elliptically distorted antiferromagnetic cycloid with magnetic moments that feature finite projections along $\pm\left[001\right]$ and $\pm\left[110\right]$. The amplitudes along these two directions exhibit a ratio of 1.58 : 1. The modulation length of the staggered magnetization amounts to $c/\left(2\delta_{1}\right)\approx 7\cdot c \approx 67 \,\mathrm{\AA}$ along $\left[001\right]$ and $a/\epsilon\cdot \sqrt{2}/4\approx 36\cdot a\sqrt{2} \approx 215\,\mathrm{\AA}$ along $\left[110\right]$ with an incommensurate parameter given by $\delta_1:=\delta_{\mathrm{cycl}}\approx\delta_{\mathrm{con}}\approx0.073(6)$. On crystallographic scales, the magnetic structure essentially displays A-type antiferromagnetic order. Accordingly, the lattice may be decomposed in two primitive sublattices, on both of which the structure represents a ferromagnetic cycloid with modulation lengths that are twice as large as the modulation lengths of the staggered magnetization.

In finite field, the magnetic structures comprise of an antiferromagnetic modulation and an uniform magnetization along the field direction. In this context, the expressions 'cycloidal' and 'conical' were used as follows. Only at zero magnetic field, the magnetic ground state represents an undistorted planar antiferromagnetic cycloid with compensated net magnetization. In finite field, the cycloid is distorted. For the domains with in-plane modulation perpendicular to the field, the cycloid is deformed into a noncoplanar cone, whereas for domains with in-plane modulation along the field, the cycloid remains coplanar but is distorted within the plane of spin rotation. For sufficiently small magnetic fields, these distortion remain comparably weak and the corresponding regime in the magnetic phase diagram is referred to as ‘cycloidal’. With further increasing magnetic field, only the noncoplanar domains survive, whereas the other domains are depopulated. The corresponding regime in the magnetic phase diagram is referred to as ‘conical’.

In the conical phase, the antiferromagnetic modulation represents an elliptically distorted cycloid with moments along $\pm\left[001\right]$ and $\pm\left[110\right]$. The amplitudes along these two axes exhibit a ratio of 1.73 : 1. The modulation lengths are essentially equal to those in cycloidal phase. The superposition of the antiferromagnetic cycloid with uniform magnetization results in the conical ground state.

In the commensurate phase, the antiferromagnetic modulation represents an A-type antiferromagnet with projections of moments along $\pm\left[001\right]$ having an uniform staggered magnetization. The magnetic structure, which represents a superposition of this antiferromagnetic modulation together with uniform magnetization for that the moments point along $\left[\bar{1}10\right]$, is a coplanar, noncollinear antiferromagnet.

In the fan-like phase, the antiferromagnetic modulation represents an amplitude modulation that has projections of magnetic moments along $\pm\left[001\right]$. The staggered magnetization has a modulation length $c/\left(2\delta_{2}\right)=4\cdot c \approx 79\,\mathrm{\AA}$ along $\left[001\right]$ with the incommensurate parameter $\delta_2:=\delta_{\mathrm{fan}}=0.113(6)$, whereas perpendicular to $\left[001\right]$ it remains constant. The magnetic structure, which is a superposition of this amplitude modulation and uniform magnetization with moments pointing along $\left[\bar{1}10\right]$, represents a coplanar antiferromagnet that is noncollinear.

\subsection{Canting axis and canting angle}

The local noncollinear canting of the two antiferromagnetic sublattices may be described quantitatively by means of a canting angle and a canting axis. For a rigorous definition of these quantities, the two antiferromagnetic sublattices $S_a$ and $S_b$ are considered as connected by a primitive translation along the bond $[111]$. When choosing two magnetic moments $\bm{M}_a$ and $\bm{M}_b$ from the sublattices $S_a$ and $S_b$ such that their lattice sites are connected by the next-nearest-neighbor bond $[111]$, the canting axis is given by $(\bm{M}_a\times\bm{M}_b)/(|\bm{M}_a\times\bm{M}_b|)$ and the canting angle is given by $\arcsin(|\bm{M}_a\times\bm{M}_b|/(M_a\cdot M_b))$. Therefore, the canting angle is largest when each sublattice encloses an angle of 45~deg with the field direction, i.e., $[\bar{1}10]$.

In the cycloidal phase at zero magnetic field, the canting axis is spatially homogeneous, whereas the canting angle varies as a function of the real-space coordinates. In the conical phase at finite magnetic field, both the canting axis and angle vary as a function of the real-space coordinates. Specifically, in zero magnetic field for each domain the magnetic structure is coplanar and the canting axis is perpendicular to the plane in which the spin cycloid rotates. In finite field, the canting axis forms a noncoplanar vector field. In both the cycloidal and the conical phase, the modulation period of spatially inhomogeneous quantities is commensurate with the magnetic modulation length.

In the commensurate phase, the canting axis and angle are spatially homogeneous. The canting axis is parallel to $[110]$ and therefore perpendicular to the magnetic field direction $[\bar{1}10]$. As a function of increasing field, the angle between each sublattice and the magnetic field direction continuously decreases, as reflected by the size of the net magnetization. When using the measured magnetization as a probe for this angle, at low temperatures the maximum canting angle is expected around 4.5~T, i.e., well within the commensurate phase~\cite{2022_Bauer_PhysRevMaterials}. Therefore, as a function of increasing field, in the commensurate phase the canting angle initially increases before decreasing again for fields exceeding 4.5~T.

In the fan-like phase, the canting axis and angle depend on the real-space coordinates. The axis is directed along $[110]$ and $[\bar{1}\bar{1}0]$ in the two halves of the magnetic unit cell, respectively. The canting angle is homogeneous in the tetragonal basal plane, but oscillates along the $[001]$ direction.

\subsection{Nonmagnetic scattering contributions}
\begin{figure}
	\includegraphics{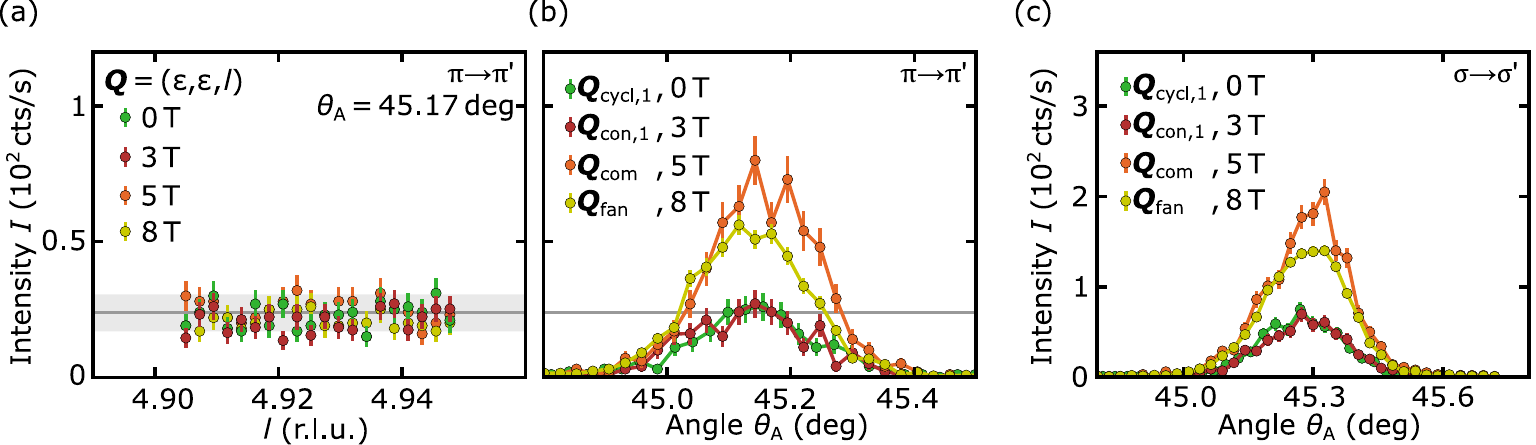}
	\caption{\label{figureS11B}Charge-scattering peaks in the fan-like and commensurate phases. (a) Diffuse charge-scattering in the $\pi\rightarrow\pi'$ channel as a function of field. The intensity is independent of $\bm{Q}$ at 22(7) cts/s (gray horizontal line with the shading indicating the standard deviation).	(b) Analyser rocking-scans in each of the magnetically long-range ordered phases for the $\pi\rightarrow\pi'$ channel revealing considerable intensity in the the commensurate and fan-like phases that emerges in addition to the diffuse scattering signal. (c) Analyser rocking-scans in each of the magnetically long-range ordered phases for the $\sigma\rightarrow\sigma'$ channel.}
\end{figure}

For each of the FLPAs presented above the structure refinements resulted in significantly better fits, when nonmagnetic scattering contributions as described by charge scattering were included in the scattering amplitude. To some extent, these contributions may account for diffuse charge-scattering that was observed in the vicinity of magnetic Bragg peaks in the $\pi\rightarrow\pi'$ channel in terms of intensity that is relatively constant in $\bm{Q}$ (see Fig.~\ref{figureS3}). This diffuse scattering signal is independent of magnetic field, as shown in Fig.~\ref{figureS11B} (a) for the $(001)$-axis and reflected by a uniform intensity of around 24(7) cts/s in the $\pi\rightarrow\pi'$ channel, which is equal within the errorbars to the diffuse scattering intensity observed in Fig.~\ref{figureS3}. 

However, as shown in Fig.~\ref{figureS11B} (b) additional charge-scattering emerges in the $\pi\rightarrow\pi'$ channel on the magnetic Bragg peaks of the commensurate and fan-like phases in addition to the diffuse scattering signal. Scattering in this channel may in principle be magnetic or non-magnetic. However, for all magnetic ground states in EuPtSi$_3$ and the $\bm{Q}$-positions in Fig.~\ref{figureS11B} (b), the moments corresponding to antiferromagnetic modulations lie strictly in the scattering plane, which implies $z_2=0$ in Eq.~\ref{equation:resonantdipolescatteringamplitude} and hence vanishing magnetic intensity in the $\pi\rightarrow\pi'$ channel. Therefore intensity presented in Fig.~\ref{figureS11B} (b) is of non-magnetic origin. 

The field dependence of the charge signal was studied also in the $\sigma\rightarrow\sigma'$ channel, in which all scattering is purely non-magnetic. Compared to the $\pi\rightarrow\pi'$ channel the scattering amplitude is enhanced by a factor $\cos^{-2}(2\theta)$ (cf. Eq.~\ref{equation:resonantdipolescatteringamplitude}).

Tab.~\ref{tableS3b} gives an overview of all reciprocal space positions, where charge scattering was observed. On an absolute scale the strength of all charge contributions is reflected by intensity in the $\pi\rightarrow\pi'$ channel as well as by the integrated intensity in the $\sigma\rightarrow\sigma'$ channel. The $D_1$ parameter indicates the strength of charge-scattering relative to the magnetic contributions in the total scattering amplitude.

\begin{table}
	\caption{\label{tableS3b}Charge-scattering in the long-range ordered phases of EuPtSi$_3$. The table summarizes all reciprocal space positions, where signatures of charge scattering were observed. The third and fourth column show integrated intensity of the analyzer rocking scan in the $\sigma\rightarrow\sigma'$ channel and the maximum intensity observed in the $\pi\rightarrow\pi'$ channel at the respective $\bm{Q}$-position. For all peak positions at that a FLPA was carried out the goodness of fit observed for the magnetic ground states regarding pure magnetic scattering (m) and with charge scattering contributions (m+c), respectively, is presented. The last column presents the charge-parameter relative to the magnetic parameter in the structure refinements.}
	\begin{ruledtabular}
		\begin{tabular}{c|c|c|c|c|c|c}
Magnetic Phase	&	$\bm{Q}$	&$\sigma\rightarrow\sigma'$ & $\pi\rightarrow\pi'$             &  $\chi^2_{\nu}$  &  $\chi^2_{\nu}$ & $D_1$\\
                &               &     int. Int. (au)           &   max. Int. (cts/s)     &    (m)        &      (m+c)                  & \\ \hline
Cycloid& $\boldsymbol{Q}_{\mathrm{cycl},1}$ &33.8(1)	& 26(5) &	24&	13& -0.04\\
& $\boldsymbol{Q}_{\mathrm{charge}}$  &  	& 81(3) &	 &	 &\\ \hline
Cone& $\boldsymbol{Q}_{\mathrm{con},1}$ & 32.8(1)&   27(5)	  &	20&	16& -0.02\\
 &  $\boldsymbol{Q}_{\mathrm{charge}}$  &  	& 77(6) &	 &	& \\ \hline
Commensurate&   $\boldsymbol{Q}_{\mathrm{com}}$ &89.7(1)  &80(9)	&	27&	4&  -0.15  \\ \hline
Fan&  $\boldsymbol{Q}_{\mathrm{fan}}$  &74.2(1) &  56(4)	&2273 &	9& -0.93
		\end{tabular}
	\end{ruledtabular}
\end{table}

\clearpage

\bibliography{EuPtSi3NoncollinearAFMSupplement}